\documentclass[american,aps,pra,reprint,groupedaddress,twocolumn,showpacs]{revtex4-2} 
\usepackage[unicode=true,pdfusetitle, bookmarks=true,bookmarksnumbered=false,bookmarksopen=false, breaklinks=false,pdfborder={0 0 0},backref=false,colorlinks=false] {hyperref}
\hypersetup{ colorlinks,linkcolor=myurlcolor,citecolor=myurlcolor,urlcolor=myurlcolor}
\usepackage{braket,colortbl,amsthm,amsmath,amssymb,txfonts}
\definecolor{myurlcolor}{rgb}{0,0,0.7}
\usepackage{graphics,graphicx}
\usepackage{xcolor}
\usepackage{color}
\usepackage{colortbl}
\usepackage{subfigure}
\usepackage{bbm}
\usepackage{dsfont}
\usepackage{float}
\usepackage{hyperref}
\usepackage[font=small,labelfont=bf,
justification= centerlast,
format=hang]{caption}
\newcommand{\tr}{\textcolor{red}}

\usepackage{ragged2e}
\usepackage{color}
\usepackage{braket}
\usepackage{physics}
\usepackage{tikz}
\usepackage{color}
\usepackage{braket}
\usepackage{physics}
\usepackage{tikz}
\usetikzlibrary{shapes.geometric, arrows}
\usepackage{latexsym}
\usepackage{bm}
\usepackage{graphics,epstopdf}
\usepackage{color}
\usepackage[usenames,dvipsnames,svgnames]{pstricks}
\usepackage{epsfig}
\usepackage{pst-grad} 
\usepackage{pst-plot} 
\usepackage[mathscr]{euscript}

\begin{document}
\title{Negativity of Wigner distribution function as a measure of incompatibility}
\author{Jatin Ghai$^{1,2}$}
\email{jghai@imsc.res.in}
\author{Gautam Sharma$^{1,2,3}$}
\email{gautam.oct@gmail.com}
\author{Sibasish Ghosh$^{1,2}$}
\email{sibasish@imsc.res.in}
\affiliation{$^1$Optics and Quantum Information Group, The Institute of Mathematical Sciences, C. I. T. Campus, Taramani, Chennai 600113, India}
\affiliation{$^2$Homi Bhabha National Institute, Training School Complex, Anushaktinagar, Mumbai 400094, India}
\affiliation{$^3$Center for Theoretical Physics, Polish Academy of Sciences, Aleja Lotników 32/46, 02-668 Warsaw, Poland}
\begin{abstract}
Measurement incompatibility and the negativity of quasiprobability distribution functions are well-known non-classical aspects of quantum systems. Both of them are widely accepted resources in quantum information processing. We acquaint an approach to establish a connection between the negativity of the Wigner function, a well-known phase-space quasiprobability distribution, of finite-dimensional Hermitian operators and incompatibility among them. We calculate the negativity of the Wigner distribution function for noisy eigenprojectors of qubit Pauli operators as a function of the noise and observe that the amount of negativity increases with the decrease in noise vis-à-vis the increase in the incompatibility. It becomes maximum for the set of maximally unbiased operators. Our results, although qualitatively, provide a direct comparison between relative degrees of incompatibility among a set of operators for different amounts of noise. We generalize our treatment for higher dimensional qudits for specific finite-dimensional Gell-Mann operators to observe that with an increase in the dimension of the operators, the negativity of their Wigner distribution, and hence incompatibility, decreases.
\end{abstract}
\maketitle

\section{Introduction}
Measurement incompatibility, i.e., the existence of measurements that can not be performed simultaneously, is one of the  features of quantum theory that demarcates it from classical physics. The earliest mention of incompatibility can be traced back to Heisenberg's uncertainty principle \cite{heisenberg} and Bohr's complementarity principle \cite{bohr1928quantum}. Although the notion of incompatibility seems like a restriction on quantum measurements it has been found to be the resource  necessary for observing quantum phenomena like Bell Non-locality \cite{PhysRevLett.48.291, PhysRevA.73.012112, PhysRevLett.103.230402}, Quantum Steering \cite{PhysRevLett.113.160402,PhysRevLett.113.160403,PhysRevLett.115.230402}, Quantum Contextuality \cite{kochen1975problem,PhysRevA.99.020103}, etc. These quantum phenomena and hence, incompatibility power many quantum information processing tasks like quantum cryptography, quantum computation, quantum state discrimination, and quantum communication. As the notion of incompatibility is crucial both fundamentally and also in applications of quantum theory, it is of interest to understand what aspects of non-classicality it can capture.

On the other hand, it has also been argued that the negativity of joint probability distributions is an indicator of non-classicality \cite{kenfack2004negativity}. One of the reasoning behind this is based on the fact that the Wigner distribution function becomes non-negative only for the coherent or the squeezed vacuum states from Hudson's theorem \cite{HUDSON1974249}. In fact, Hudson's theorem has been shown to hold also for discrete Wigner distributions \cite{gross2006hudson}. Thus, to some extent, one can say that the negativity of joint probability distributions is a good quantifier of non-classicality. 

Later, it was also shown in  \cite{PhysRevLett.101.020401}, that the notion of negativity of joint probability distributions is equivalent to contextuality. Inspired by the fact that measurement incompatibility is also necessary for quantum contextuality, we are interested in identifying how measurement incompatibility leads to negative joint probability distributions. Such a query has recently been explored in \cite{PhysRevA.104.042212}, where, by using $\textbf{S}$-ordered phase space distributions it was shown that the non-negativity of the joint probability distribution of POVM (positive operator-valued measurement) elements is a sufficient condition for their joint measurability. They applied their approach to analyze incompatibility-breaking channels and derived sufficient conditions for joint measurability for bosonic systems and Gaussian channels. 

On the other hand, in this work, we try to use the negativity of the generalized Wigner distribution function introduced in \cite{schwonnek2020wigner} as an indicator, although qualitative, of the degree of incompatibility. This Wigner distribution function is generalized in the sense that it is defined for arbitrary Hermitian operators. Though it is implicit in the definition of incompatibility of POVMs, the incompatibility must come from the non-existence of a joint probability distribution. However, it is desirable that this joint probability distribution comes from a Wigner-like quasi-probability distribution. We are able to achieve this partially such that we are able to show qualitatively for a special set of measurements the relation between the degree of incompatibility and the negative volume of the Wigner distribution by introducing a tunable noise parameter. Moreover, for more than two PVMs (projection-valued measurement), a quantifier has been missing so far\cite{Heinosaari_2016}. In this work, we are also able to show that the negativity of the Wigner distribution function can be a good candidate for the indicator of incompatibility among more than three PVMs.

This paper is organized as follows: Sec.(\ref{Sec2}) encompasses the definition and some basic properties of the generalized Wigner function. In Sec.(\ref{Sec4}) we calculate the Wigner function for the noisy eigenprojectors of these qubit Pauli operators. We know that the addition of noise leads to a decrease in incompatibility. Thus, we calculate the negativity of the Wigner function as a function of the noise parameter to try to establish a connection between both. In the succeeding Sec.(\ref{Sec5}) we do a similar treatment for some specific noisy qubit Pauli operators. In Sec.(\ref{Sec7}) we generalize this procedure to arbitrary dimensional qudits for Gell-Mann operators of arbitrary dimension having some specific form. Sec.(\ref{Sec8}) concludes this paper and provides an outlook for future research directions.

\section{Wigner distribution function for n arbitrary observables}\label{Sec2}
We begin by introducing the generalization of the Wigner distribution function over a set of arbitrary Hermitian operators. We will briefly present the definition and properties of the Wigner distribution function here. For more details, we request the reader to refer to the original paper \cite{schwonnek2020wigner}.

To define such a Wigner distribution function, we consider a set of $n$ bounded Hermitian operators $\{A_1, A_2,..., A_n\}$ on a $d$-dimensional Hilbert space. One of the basic properties of a Wigner distribution function is that it should reproduce the correct marginals of all possible linear combinations  $\vec{\xi}\cdot \vec{A}=\sum_k\xi_kA_k$, of the operators, i.e., the Wigner distribution function must satisfy the following condition
\begin{align}\label{margcond}
	\int d^{n} a \mathscr{W}_{\rho}\left(a_{1}, \ldots, a_{n}\right) f(\vec{\xi} \cdot \vec{a}) 
	={\rm tr} [{\rho}f({\vec{\xi}}.\vec{A})],
\end{align}
where $f:\mathbb{R}\rightarrow\mathbb{C}$ is a bounded infinitely differentiable function.

\textit{\textbf{Definition 1:}} If we take test function $f(t)=e^{-it}$ in Eqn. (\ref{margcond}), we can get the expression of the Fourier transform of $\mathscr{W}_{\rho}(\vec{a})$ as
\begin{align}\label{fourierTrans}
	    \widehat{\mathscr{W}}_{\rho}(\vec{\xi})={\rm tr} [\rho e^{i \vec{\xi} \cdot \vec{A}}].
\end{align}
Using the above definition, the Wigner distribution function can be defined as the reverse Fourier transform
\begin{align}\label{wignerdist}
	\mathscr{W}_{\rho}(\vec{a})=\frac{1}{(2 \pi)^{n}} \int d^n  \xi e^{-i \vec{\xi} \cdot \vec{a}} \widehat{\mathscr{W}}_{\rho}(\vec{\xi}).
\end{align}
It can be checked that this expression satisfies the marginal property \textit{i.e.} if we integrate over one of the phase space parameters we will retrieve a valid phase space distribution for the remaining parameters. Hence can be used as a Wigner distribution function.

There is another way to define the Wigner distribution function using the "Weyl-ordered moments" which also satisfies Eq.(\ref{margcond}) as shown in \cite{schwonnek2020wigner}, but we will skip it here as it is not needed for our purpose.
\subsection{Graphical Representation}As we will see in the later sections, the Wigner distribution function in Eq. (\ref{wignerdist}) may not be well-defined at every point in the phase space. In order to visualize the behavior of such a distribution, we regularize the Wigner distribution function by taking its convolution with a Gaussian $G_{\varepsilon}$ peaked at the origin and having covariance $\varepsilon$. This can be done by multiplying \ref{wignerdist} with $\widehat{G}_{\varepsilon}$ (inverse Fourier transform of $G_{\varepsilon}$ which is again a Gaussian) and taking the Fourier transform,  i.e., 
\begin{align}\label{graphical}
	\mathscr{W}*G_{\varepsilon}=\frac{1}{(2 \pi)^{n}} \int d^n \xi e^{-i \vec{\xi} \cdot \vec{a}} \widehat{\mathscr{W}}_{\rho}(\vec{\xi})\widehat{G}_{\varepsilon},
\end{align}
 The regularized Wigner distribution function might not give the correct marginal distributions but it will help in the visualization of the Wigner distribution function properties. We see that $\mathscr{W}_{\rho}(a)=\lim_{\varepsilon\rightarrow0}\mathscr{W}_{\rho}(a)*G_{\varepsilon}$ is an alternative definition of $\mathscr{W}_{\rho}(a)$.

\subsection{Basic Properties}
The generalized Wigner distribution function has several properties which are markedly different from the well-known form of the Wigner distribution function. We list only a few which are relevant to us. 
\begin{enumerate}
	\item $\mathscr{W}_{\rho}(\vec{a})$ is a real valued distribution. This follows from Definition 1 and by noting that $\overline{\widehat{\mathscr{W}}_{\rho}(\vec{\xi})}=\widehat{\mathscr{W}}_{\rho}(-\vec{\xi})$.
	\item The Wigner distribution function has support in a compact convex set, which is equivalent to the joint numerical range of the operators $A_k$ over all the density matrix operators, i.e., 
	\begin{align*}
		\mathscr{W}_{supp}=\mathcal{R}=\left\{\{a_1,a_2,...,a_n\} \in \mathbb{R}^{n} \mid a_{k}={\rm tr} [\rho A_{k}]\right\}.
	\end{align*}
	This means that $\forall \rho, \mathscr{W}_{\rho}(\vec{a})=0$ outside of $\mathcal{R}$.
	\item One of the most distinguishing features of $\mathscr{W}_{\rho}(\vec{a})$ are its singularities. For finite-dimensional matrices $A_k$, the singularities of  $\mathscr{W}_{\rho}(\vec{a})$ lie  on the closure of following set 
	\begin{align*}
		 \mathcal{S}=\left\{\vec{a} \in \mathbb{R}^{n} \mid a_{k}=\left\langle\psi\left|A_{k}\right| \psi\right\rangle\right.\\
		\|\psi\|=1, (\vec{\xi} \cdot \vec{A}) \psi=\lambda \psi\}, 
	\end{align*}
where $\psi$ can be any eigenvector corresponding to a non-degenerate eigenvalue of any $\vec{\xi}\cdot \vec{A}$. $\lambda$ are those non-degenerate eigenvalues.

As we will see in the later sections, for certain examples, the Wigner distribution function indeed takes the form of a delta function. To visualize such a function, we will use the graphical representation of the Wigner distribution function.
\item When $A_k$ are finite dimensional matrices and $\rho$ is a full-rank density matrix operators, $\mathscr{W}_{\rho}(\vec{a})$ is positive if and only if all the $A_k$s  commute with each other. In such a case $\mathscr{W}_{\rho}(\vec{a})$ is a sum of $\delta$-functions with weights dependent on $\rho$
\end{enumerate}
For the proofs of properties 2, 3 and 4, we refer the reader to \cite{schwonnek2020wigner}. Property 4 motivates us to investigate how the negativity of the Wigner distribution function depends on the Incompatibility(or commutativity of sharp operators) of Hermitian operators $A_k$.

\section{Wigner distribution for noisy projections of eigenvectors of Pauli operators}\label{Sec4}
In this section, we will explicitly calculate the Wigner function for noisy projections of eigenvectors of Pauli vectors. The procedure is on the same lines as for the maximally unbiased case (noiseless Pauli operators) in \cite{schwonnek2020wigner}. We know that the noiseless eigenprojections are maximally incompatible with each other. Adding noise to them in the form of identity will decrease this incompatibility. We will then calculate the negative volume of the Wigner distribution function for different choices of the noise parameter. First, let us consider two of these noisy projections.
\subsection{Two operators (n=2)}
\subsubsection{Calculating the Wigner function}
We will consider the qubit operators to be $A_k=\frac{\lambda}{2}I+(1-\lambda)\vert + \rangle_k \langle + \vert, k=1,2$ where $\{\vert + \rangle_1\langle + \vert, \vert + \rangle_2\langle + \vert\}$ are respectively the projections of the eigenvectors of Pauli matrices $\{\sigma_x,\sigma_y\}$ having eigenvalue +1 and $0\leq\lambda\leq1$ is the noise parameter. We note that the following relation holds irrespective of the choice of operators
\begin{align}\label{derivativeofWhat}
	\frac{\partial}{\partial \xi_{k}} \widehat{\mathscr{W}}_{\mathbb{I}}(\vec{a})=i \tr [A_{k} e^{i \vec{\xi} \cdot \vec{A}}].
\end{align}
which follows from the definition of $\widehat{\mathscr{W}}_{\rho}(\vec{a})$. For a qubit density matrix $\rho=(\mathbb{I}+\sum_kr_k\sigma_k)/2$, where $\vec{r}=\{r_1,r_2\}$ is the Bloch vector, the Wigner function can be written as
\begin{align}\label{n2projidentorhoproj}
	\mathscr{W}_{\rho}^{proj++}(\vec{a})=\frac{1}{2}(1+\vec{r} \cdot \vec{a'})\mathscr{W}_{\mathbb{I}}^{proj++}(\vec{a}),
\end{align}
where
\begin{align}
    a'_1=\frac{2a_1-1}{(1-\lambda)},\nonumber\\
    a'_2=\frac{2a_2-1}{(1-\lambda)}.
\end{align}
The above relation is obtained by doing integration by parts in Eqn.(\ref{wignerdist}) and then using Eqn.(\ref{derivativeofWhat}). Moreover, this equation is always satisfied whenever the state $\rho$ can be written as the linear combination of the operators $A_k$. The Fourier transform of the Wigner function $\hat{\mathscr{W}}_{\mathbb{I}}^{proj++}(\vec{a})$ comes out to be $2e^{i\frac{(\xi_{1}+\xi_{2})}{2}}\cos(\frac{(\lambda-1}{2}|\xi|)$ where $|\xi|=\sqrt{\xi^{2}_{1}+\xi^{2}_{2}}$. By convoluting it with a scaled function of the form $e^{-\frac{\varepsilon(1-\lambda)|\xi|}{2}}$. The Wigner function comes out to be
\begin{align}
        \mathscr{W}_{\mathbb{I}}^{proj++}(\vec{a})&= \frac{1}{(2 \pi)^{2}} \int d^2 \xi e^{-i\vec{\xi} \cdot \vec{a}}2e^{i\frac{(\xi_{1}+\xi_{2})}{2}}\cos(\frac{(\lambda-1}{2}|\xi|)e^{-\frac{\varepsilon(1-\lambda)|\xi|}{2}}\nonumber\\
        &= \frac{4}{(1-\lambda)^2(2 \pi)^{2}} \int d^2\xi' e^{-i\vec{\xi}' \cdot \vec{a}'}2\cos{|\xi'|}e^{-\varepsilon |\xi'|}\nonumber\\
       &=\frac{-4}{(1-\lambda)^2\pi\left(1-|a'|^{2}\right)^{3 / 2}},
\end{align}
where we have used the transformation $\vec{\xi}'=\frac{(1-\lambda)\vec{\xi}}{2}$. Using Eqn.(\ref{n2projidentorhoproj}) the total Wigner function for state $\rho$ is given as
\begin{equation}\label{wigproj2}
\mathscr{W}_{\rho}^{proj++}(\vec{a})=-\frac{(1+\vec{r} \cdot \vec{a}')}{2\pi(1-\lambda)^2}\frac{4}{\left(1-|a'|^{2}\right)^{3 / 2}},
\end{equation}
But this is not the only combination possible for the choice of noisy projections of eigenvectors of Pauli operators. In fact there are three more possibilities. Namely
\begin{itemize}
    \item $\{(\frac{\lambda}{2}I+(1-\lambda)\vert + \rangle_1 \langle + \vert),(\frac{\lambda}{2}I+(1-\lambda)\vert - \rangle_2 \langle - \vert)\} $
    \item $\{(\frac{\lambda}{2}I+(1-\lambda)\vert - \rangle_1 \langle - \vert),(\frac{\lambda}{2}I+(1-\lambda)\vert + \rangle_2 \langle + \vert)\} $
    \item $\{(\frac{\lambda}{2}I+(1-\lambda)\vert - \rangle_1 \langle - \vert),(\frac{\lambda}{2}I+(1-\lambda)\vert - \rangle_2 \langle - \vert)\} $
\end{itemize}
The calculations for these three proceeds in the same way as in the previous case. The expression for the Wigner function comes out to be
\begin{align}
    &\mathscr{W}_{\rho}^{proj+-}(\vec{a})=-\frac{(1+r_1 a'_1-r_2 a'_2)}{2\pi(1-\lambda)^2}\frac{4}{\left(1-|a'|^{2}\right)^{3 / 2}},\label{wigproj+-}\\
    &\mathscr{W}_{\rho}^{proj-+}(\vec{a})=-\frac{(1-r_1 a'_1+r_2 a'_2)}{2\pi(1-\lambda)^2}\frac{4}{\left(1-|a'|^{2}\right)^{3 / 2}},\label{wigproj-+}\\
    &\mathscr{W}_{\rho}^{proj--}(\vec{a})=-\frac{(1-\vec{r}\cdot \vec{a}')}{2\pi(1-\lambda)^2}\frac{4}{\left(1-|a'|^{2}\right)^{3 / 2}},\label{wigproj--}
\end{align}
\subsubsection{Comparing the negativity}
Now we will calculate the negativity of the Wigner function calculated in the previous section and analyze its variation with the noise parameter $\lambda$. The negative volume is calculated by integrating it over the phase space region where its value is negative. Mathematically
\begin{align}\label{negcalproj}
	\mathscr{W}_{neg}(\rho,\vec{A})= \int\frac{1}{2}(\mathscr{W}_{\rho}(\vec{a})-|\mathscr{W}_{\rho}(\vec{a})|)d^2a,
\end{align}
where $\vec{A}=\{A_1, A_2\}$ is the operator set for which the Wigner function is calculated.

Borrowing the expression of the Wigner function for the MUB case from \cite{schwonnek2020wigner} the negativity can be calculated as:
\begin{align}\label{mubnegn2proj}
	\mathscr{W}_{neg}^{mub}(\rho,\vec{A})&= \int\frac{1}{2}\left(\mathscr{W}_{\rho}^{mub}(\vec{a})-|\mathscr{W}_{\rho}^{mub}(\vec{a})|\right)d^2a  \nonumber\\
	&= \int\frac{1}{2}\Bigg(-\frac{(1+\vec{r} \cdot \vec{a})}{2\pi}\frac{1}{\left(1-|a|^{2}\right)^{3 / 2}} \nonumber \\
	&-\Big|-\frac{(1+\vec{r} \cdot \vec{a})}{2\pi}\frac{1}{\left(1-|a|^{2}\right)^{3 / 2}}\Big|\Bigg)d^2a
\end{align}
Now we will compare the negativity of noisy cases with the MUB case. But we notice that the regularising function used in the noisy cases is different from what was used in that. Thus, in order to compare them we have to normalize the Wigner distribution function in Eqn.(\ref{wigproj2}) by dividing it by the factor
\begin{align*}
	\frac{I_{G'}}{I_{G}}&=\frac{\int d^2\xi e^{-\frac{\varepsilon(1-\lambda)}{2}\sqrt{ \xi^{2}_{1}+\xi^{2}_{2}} }}{\int d^2\xi e^{-\varepsilon\sqrt{\xi_1^2+\xi_2^2}}} \\
	&=\frac{4}{(1-\lambda)^2},
\end{align*}
where $I_{G'}$ is the integral of the scaled Gaussian and $I_G$ is the integral of the Gaussian function over the Fourier space. One more thing to note here is that as the regularising functions used for different values of the noise parameter $\lambda$ are different we can't claim that the numerical value of the negativity of the Wigner function quantitatively indexes the degree of incompatibility. Rather, we can only compare the negativity for various values of the noise parameter and give its variation with the degree of incompatibility qualitatively. An argument for the validity of this normalization will be given in the next section. Thus using Eqn.(\ref{wigproj2}) the negativity can be calculated as
\begin{align}\label{n2noisyprojneg1}
	\mathscr{W}_{neg}^{proj++}(\rho,\vec{A})&= \int\frac{1}{2}\left(\mathscr{W}_{\rho}^{proj++}(\vec{a})-|\mathscr{W}_{\rho}^{proj++}(\vec{a})|\right)d^2a\nonumber\\
	&=\frac{(1-\lambda)^2}{4}\mathscr{W}_{neg}^{mub}(\rho,\vec{A}),
\end{align}
Thus from the above expression, we see that the negative volume of the Wigner function for noisy projectors is less than the corresponding negative volume for MUB and further decreases with an increase in the noise. 
Using the expression for Wigner distribution in Eqn.(\ref{wigproj2}) we can numerically integrate Eqn.(\ref{n2noisyprojneg1}) and calculate the negative volume of the Wigner distribution for various values of the noise parameter. Also, we know that with an increase in the noise the incompatibility between the observables decreases. From this, we have an indication that the Wigner function, through its negativity, indeed captures the incompatibility between observables, although qualitatively. Comparing this result with \cite{PhysRevA.104.042212} we observe that  the positivity of the general quasiprobability distribution function is the sufficient condition for joint measurability. Although analytical, their formalism gives a tight upper bound on the degree of incompatibility among Gaussian measurements for incompatibility-breaking Gaussian channels. On the other hand in our results, as we will see later also, for a quite general class of observables the negativity of the Wigner function decreases monotonically with the increases in the noise parameter vis-à-vis the incompatibility. Although the result is qualitative up to some extent it clearly gives a tool to compare the degree of incompatibility among operators for different amounts of noise added.

Similarly, if we calculate the negative volume for Wigner distribution in Eqns.(\ref{wigproj+-},\ref{wigproj-+},\ref{wigproj--}) by substituting them in Eqn.(\ref{negcalproj}) we get exactly the same result as we got for Wigner distribution in Eqn.(\ref{wigproj2}) in Eqn.(\ref{n2noisyprojneg1}). Hence for further calculation, we will only calculate and analyze the Wigner function for the noisy projectors having eigenvalue +1 only. One important thing to note is that the negative volume of the Wigner function doesn't depend on the Bloch vector $\vec{r}=(r_1,r_2)$.
\subsection{Three operators (n=3)}

\subsubsection{Calculating the Wigner distribution}
Now we will calculate the Wigner function for noisy projections of all three Pauli matrices. Qubit operators are $A_k=\frac{\lambda}{2}I+(1-\lambda)\vert + \rangle_k \langle + \vert, k=1,2,3$ where $\{\vert + \rangle_1\langle + \vert, \vert + \rangle_2\langle + \vert, \vert + \rangle_3\langle + \vert\}$ are respectively the projections of the eigenvectors of Pauli matrices $\{\sigma_x,\sigma_y,\sigma_z\}$ with eigenvalue +1 and $\lambda$ is the noise parameter. For a general qubit density matrix  $\rho=(\mathbb{I}+\sum_kr_k\sigma_k)/2$, where $\vec{r}=\{r_1,r_2,r_3\}$ is the Bloch vector, the Wigner distribution function, using Eqn.(\ref{derivativeofWhat}), can be written as
\begin{align}\label{n3projidentorhoproj}
	\mathscr{W}_{\rho}^{proj}(\vec{a})=\frac{1}{2}(1+\vec{r} \cdot \vec{a}')\mathscr{W}_{\mathbb{I}}^{proj}(\vec{a}),
\end{align}
where
\begin{align}
    a'_1=\frac{2a_1-1}{(1-\lambda)},\nonumber\\
    a'_2=\frac{2a_2-1}{(1-\lambda)},\nonumber\\
    a'_3=\frac{2a_3-1}{(1-\lambda)}.
\end{align}
The Fourier transform of the Wigner function $\hat{\mathscr{W}}_{\mathbb{I}}^{proj}(\vec{a})$ comes out to be $2e^{i\frac{(\xi_{1}+\xi_{2}+\xi_{3})}{2}}\cos(\frac{(\lambda-1}{2}|\xi|)$ where $|\xi|=\sqrt{\xi^{2}_{1}+\xi^{2}_{2}+\xi^{2}_{3}}$. By convoluting it with a scaled Gaussian function of the form $e^{-\frac{\varepsilon(1-\lambda)^2|\xi|^2}{4}}$. Using the transformation $\vec{\xi}'=\frac{(1-\lambda)\vec{\xi}}{2}$ the Wigner function can be calculated as
\begin{align}\label{n3noisyprojidenwigner}
	 \mathscr{W}_{\mathbb{I}}^{proj}(\vec{a})&=\frac{1}{(2 \pi)^{3}} \int d^3 \xi e^{-i \vec{\xi} \cdot \vec{a}} 2e^{i\frac{(\xi_{1}+\xi_{2}+\xi_{3})}{2}}\cos(\frac{(\lambda-1)}{2}|\xi|)e^{-\varepsilon|\xi|'^2} \nonumber \\&=\frac{-8}{\pi |a'|(1-\lambda)^3} \frac{d}{d |a'|} \frac{e^{-\frac{(|a'|-1)^{2}}{4 \varepsilon}}}{\sqrt{4 \pi \varepsilon}}  \nonumber \\&=\frac{-8\delta^{\prime}(|a'|-1)}{2 \pi |a'|(1-\lambda)^3}.
\end{align}
Therefore, the Wigner distribution function for an arbitrary qubit state can be written as
\begin{align}\label{n3noisyprojrhowigner}
	\mathscr{W}_{\rho}^{proj}(\vec{a})=-\frac{8(1+\vec{r} \cdot \vec{a}')}{4 \pi|a'|(1-\lambda)^3} \delta^{\prime}(|a'|-1).
\end{align}
\subsubsection{Comparing the negativity}
Now we will calculate the negativity of the Wigner function in the same way as we did previously for the n=2 case. The negative volume can be calculated by integrating the Wigner function over the three-dimensional phase space region where it is negative as
\begin{align}\label{n3negcalproj}
	\mathscr{W}_{neg}(\rho,\vec{A})= \int\frac{1}{2}(\mathscr{W}_{\rho}(\vec{a})-|\mathscr{W}_{\rho}(\vec{a})|)d^3a.
\end{align}

\begin{figure}[h!]
    \centering
    \includegraphics[scale=0.5]{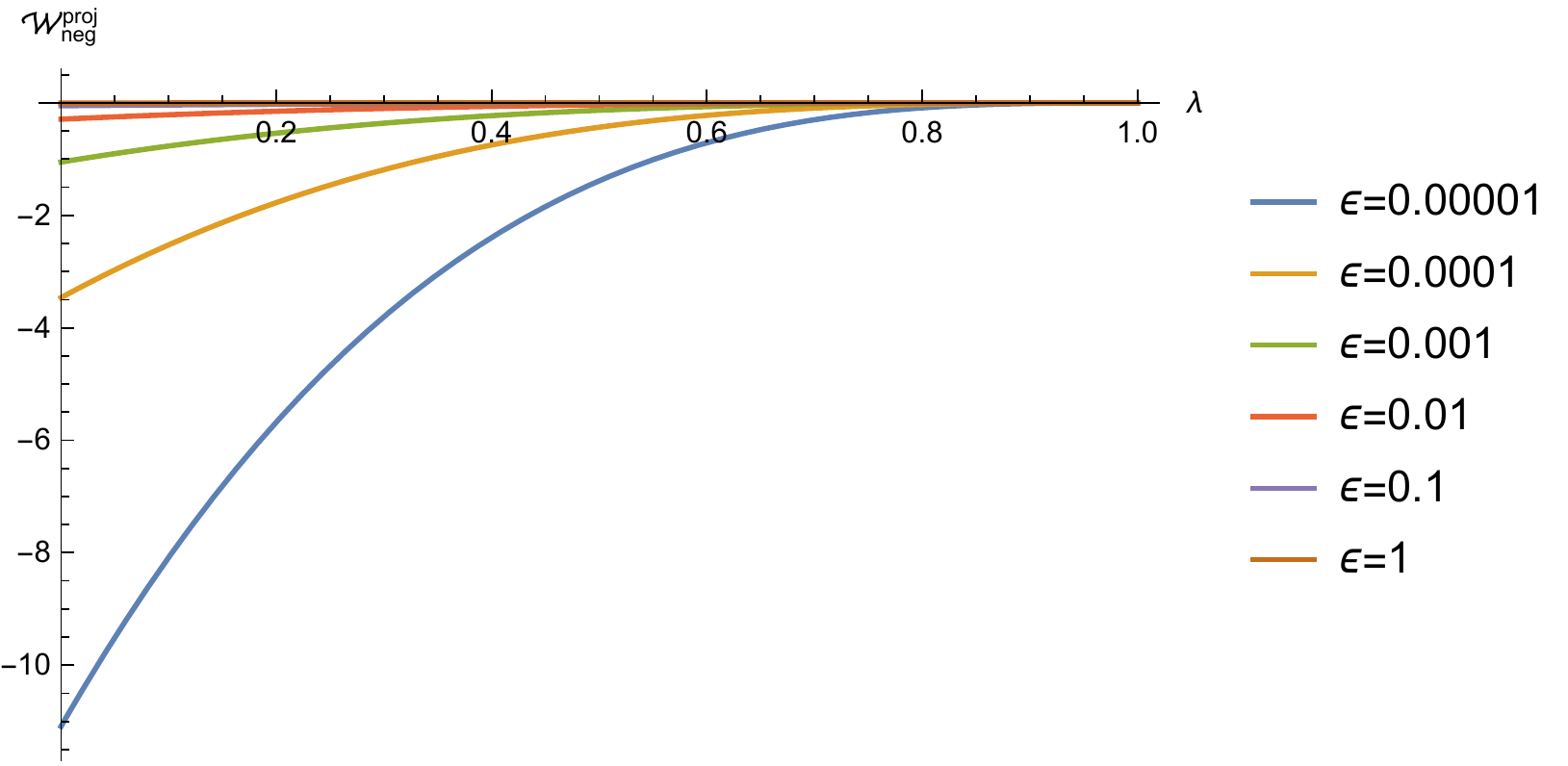}
    \caption{Variation of negativity of Wigner distribution function with noise $\lambda$ for different values of regularising parameter $\varepsilon$ under scaled Gaussian regularization}
    \label{fig:n2}
\end{figure}

Following the same logic as in the previous n=2 case in order to compare the negativity of the noisy case with the MUB case we have to normalize the Wigner distribution function in Eqn. (\ref{n3noisyprojidenwigner}) by dividing it by the factor
\begin{align*}
	\frac{I_{G'}}{I_{G}}&=\frac{\int d^3\xi e^{-\frac{\varepsilon(1-\lambda)^2}{4}(\xi^{2}_{1}+\xi^{2}_{2}+\xi^{2}_{3})} }{\int d^3\xi e^{-\varepsilon(\xi_1^2+\xi_2^2+\xi_3^2)}}, \\
	&=\frac{8}{(1-\lambda)^3},
\end{align*}
where $I_{G'}$ is the integral of the scaled Gaussian and $I_G$ is the integral of the Gaussian function over the Fourier space. Thus using Eqn. (\ref{n3noisyprojrhowigner}) the negativity can be calculated as
\begin{align}\label{n3noisyprojneg1}
	\mathscr{W}_{neg}^{proj}(\rho,\vec{A})&= \int\frac{1}{2}\left(\mathscr{W}_{\rho}^{proj}(\vec{a})-|\mathscr{W}_{\rho}^{proj}(\vec{a})|\right)d^3a\nonumber\\
	&=\frac{(1-\lambda)^3}{8}\mathscr{W}_{neg}^{mub}(\rho,\vec{A}).
\end{align}
From the above expression, we see that the negative volume of the Wigner function for noisy projectors decreases with an increase in the noise.  We calculate the negative volume of $\mathscr{W}_{\mathbb{\rho}}^{mub}(\vec{a})$ and $\mathscr{W}_{\mathbb{\rho}}^{proj}(\vec{a})$ analytically by using the Gaussian approximation of delta function. The expression in Eqn.(\ref{n3noisyprojneg1}) reduces to
\begin{equation}
        \mathscr{W}_{neg}^{proj}(\rho,\vec{A})=\frac{(\lambda-1)^{3}}{16}\left(-1+\frac{1}{\sqrt{\pi \varepsilon}}+Erfc \left(\frac{1}{2\sqrt{\varepsilon}}\right) \right).
    \end{equation}
We see from this expression that the negative volume depends on both the regularisation parameter $\varepsilon$ and the noise parameter $\lambda$. We can fix a particular value of resolution of the Wigner function by fixing $\varepsilon$ and analyze how the negative volume of the Wigner distribution decreases with an increase in the noise. The result is plotted in Fig.(\ref{fig:n2}).
We see that for every value of $\varepsilon$ negativity decreases with an increase in the noise parameter $\lambda$. Also the smaller the value of regularising parameter the better the resolution between two values of negative volume of Wigner distribution for the corresponding two values of noise parameter.

\begin{figure}[h!]
    \centering
    \includegraphics[scale=0.5]{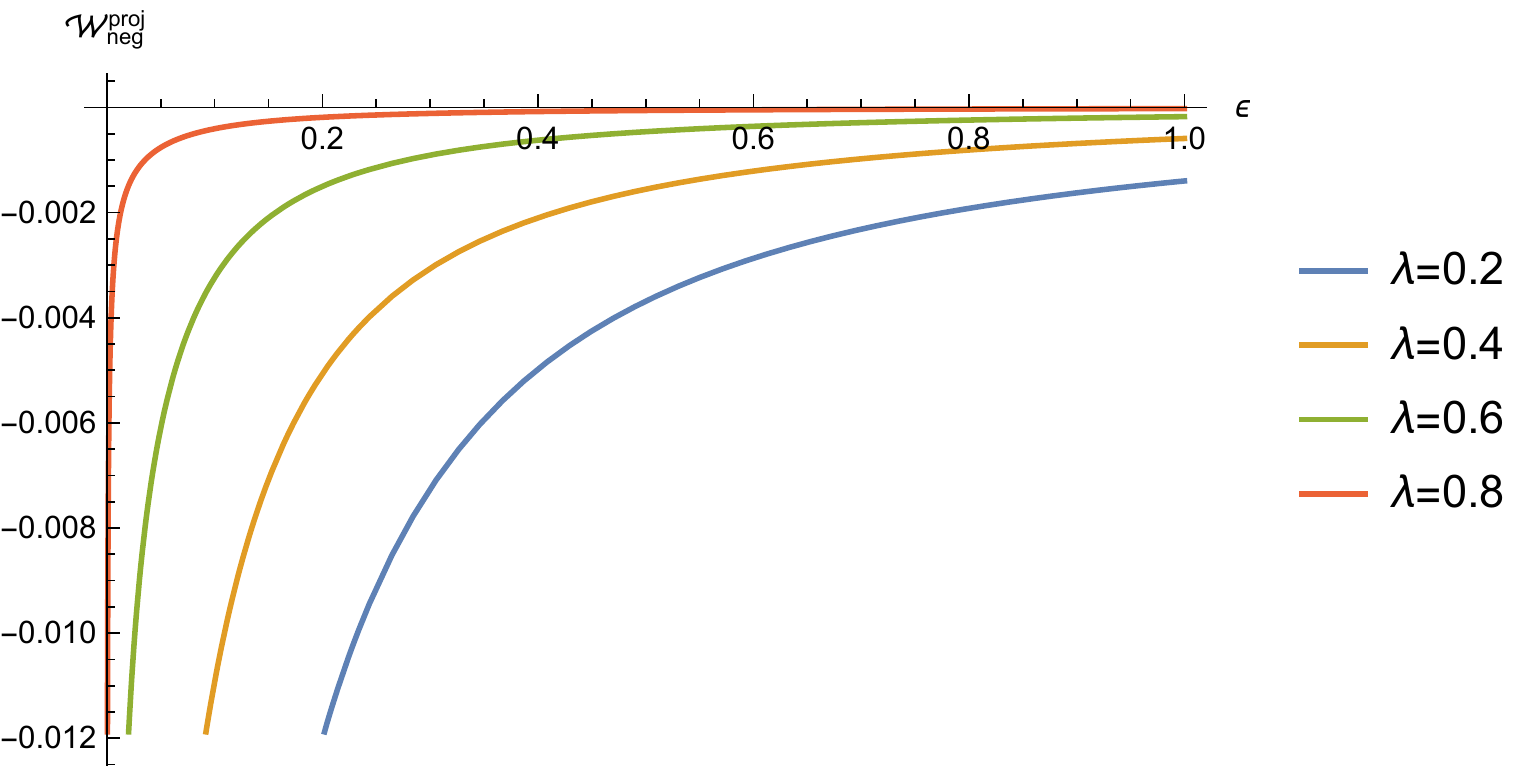}
    \caption{Variation of negativity of Wigner distribution function with regularising parameter $\varepsilon$ for different amounts of noise $\lambda$ under scaled Gaussian regularization}
    \label{fig:n3}
\end{figure}

Next, for a particular value of $\lambda$ we plot the negativity of the Wigner function against the regularizing parameter $\varepsilon$. The trend is presented in Fig.(\ref{fig:n3}). We see that this plot again reinstates the fact that as the value of regularising factor decreases the difference between the negative volume for two given values of the noise parameter keeps increasing. Hence the resolution increases. So it is preferable to have $\varepsilon \rightarrow 0$.

 We note that this is not the only way to consider n=3 noisy projections. We can consider the first two operators to be noisy projections of eigenvectors of say $\sigma_y$ corresponding to eigenvalues $\pm1$ and the third operator to be the noisy projection of eigenvector of $\sigma_x$  having eigenvalue $+1$ \textit{i.e} our operator set looks like
 \begin{itemize}
     \item $\{(\frac{\lambda}{2}I+(1-\lambda)\vert + \rangle_1 \langle + \vert),(\frac{\lambda}{2}I+(1-\lambda)\vert + \rangle_2 \langle + \vert),(\frac{\lambda}{2}I+(1-\lambda)\vert - \rangle_2 \langle - \vert)\} $
 \end{itemize}
 where $|{\pm}\rangle_1$ represent the eigenvectors of ${\sigma}_x$ while $|{\pm}\rangle_2$ represent the eigenvectors of ${\sigma}_y$ corresponding to the eigenvalues $\pm 1$'. Proceeding in the same way, we find out that for a qubit density matrix $\rho=(\mathbb{I}+\sum_kr_k\sigma_k)/2$, where $\vec{r}=\{r_1,r_2\}$ is the Bloch vector, the Wigner function can be written as
\begin{align}\label{n3projidentorhoproj2}
	\mathscr{W}_{\rho}^{proj++-}(\vec{a})=\frac{1}{2}(1+\vec{r} \cdot \vec{a}_{12}')\mathscr{W}_{\mathbb{I}}^{proj++-}(\vec{a}),
\end{align}
where $\vec{a}'_{12}=(a'_1,a'_2,0)$ and
\begin{align}
    a'_1=\frac{2a_1-1}{(1-\lambda)},\nonumber\\
    a'_2=\frac{a_2-a_3}{(1-\lambda)}.
\end{align}
The Fourier transform of the Wigner function, $\hat{\mathscr{W}_{\mathbb{I}}^{proj++-}(\vec{a})}$ comes out to be $2e^{i\frac{(\xi_{1}+\xi_{2}+\xi_{3})}{2}}\cos(\frac{(\lambda-1}{2}|\xi_{12}"|)$ where $|\xi"_{12}|=\sqrt{\xi^{2}_1+(\xi_2-\xi_3)^2}$. By convoluting it with a regularizing function of the form $e^{ (-\frac{\varepsilon(1-\lambda)|\xi"_{12}|}{2}-\frac{\varepsilon(1-\lambda)^2(\xi_2+\xi_3)^2}{4})}$. Using the transformation
\begin{align}
    \frac{(1-\lambda)}{2}\xi_1=\xi"_1,\nonumber\\
    \frac{(1-\lambda)}{2}(\xi_2-\xi_3)=\xi"_2,\nonumber\\
    \frac{(1-\lambda)}{2}(\xi_2+\xi_3)=\xi"_3.
\end{align}
The Wigner distribution can be calculated as
\begin{align}\label{n3noisyprojidenwigner2}
	 \mathscr{W}_{\mathbb{I}}^{proj++-}(\vec{a})&=\frac{1}{(2 \pi)^{3}} \int d^3 \xi e^{-i \vec{\xi} \cdot \vec{a}} 2e^{i\frac{(\xi_{1}+\xi_{2}+\xi_{3})}{2}}\nonumber \\
	&=\frac{-4}{\pi (1-\lambda)^3}\delta(a'_3)\frac{1}{(1-|a'_{12}|^2)^\frac{3}{2}}.
\end{align}
where $a'_3=\frac{a_2+a_3-1}{(1-\lambda)}$.
Thus total Wigner function can be written as
\begin{equation}\label{n3noisyprojrhowigner2}
    \mathscr{W}_{\rho}^{proj++-}(\vec{a})=\frac{1}{2}(1+\vec{r} \cdot \vec{a}_{12}')\frac{-4}{\pi (1-\lambda)^3}\delta(a'_3)\frac{1}{(1-|a'_{12}|^2)^\frac{3}{2}}.
\end{equation}

But this is not the only combination possible for choosing the three noisy eigenprojections of $\sigma_x,\sigma_y$. In fact three more combinations are possible given as
\begin{itemize}
    \item  $\{(\frac{\lambda}{2}I+(1-\lambda)\vert + \rangle_1 \langle + \vert),(\frac{\lambda}{2}I+(1-\lambda)\vert - \rangle_1 \langle - \vert)\}, (\frac{\lambda}{2}I+(1-\lambda)\vert + \rangle_2 \langle + \vert) $
    \item  $\{(\frac{\lambda}{2}I+(1-\lambda)\vert + \rangle_1 \langle + \vert),(\frac{\lambda}{2}I+(1-\lambda)\vert - \rangle_1 \langle - \vert)\}, (\frac{\lambda}{2}I+(1-\lambda)\vert - \rangle_2 \langle - \vert) $
    \item $\{(\frac{\lambda}{2}I+(1-\lambda)\vert - \rangle_1 \langle - \vert),(\frac{\lambda}{2}I+(1-\lambda)\vert + \rangle_2 \langle + \vert)\}, (\frac{\lambda}{2}I+(1-\lambda)\vert - \rangle_2 \langle - \vert) $
\end{itemize}
Following the same procedure, the Wigner function can be calculated as
\begin{align}
    &\mathscr{W}_{\rho}^{proj+-+}(\vec{a})=\frac{1}{2}(1+r_1 a'_2+r_2 a'_1)\frac{-4}{\pi (1-\lambda)^3}\delta(a'_3)\frac{1}{(1-|a'_{12}|^2)^\frac{3}{2}},\label{wigproj+-+}\\
    &\mathscr{W}_{\rho}^{proj+--}(\vec{a})=\frac{1}{2}(1+r_1 a'_2-r_2 a'_1)\frac{-4}{\pi (1-\lambda)^3}\delta(a'_3)\frac{1}{(1-|a'_{12}|^2)^\frac{3}{2}},\label{wigproj+--}\\
    &\mathscr{W}_{\rho}^{proj-+-}(\vec{a})=\frac{1}{2}(1-r_1 a'_1+r_2 a'_2)\frac{-4}{\pi (1-\lambda)^3}\delta(a'_3)\frac{1}{(1-|a'_{12}|^2)^\frac{3}{2}},\label{wigproj-+-}
\end{align}
The negativity of the Wigner distribution for these operator sets can be calculated in the same way as the previous case by using Eqn. (\ref{n3negcalproj}). But first, we have 
 to normalize the Wigner function expressions by dividing them with the factor
\begin{align*}
	\frac{I_{G'}}{I_{G}}&=\frac{\int d^3\xi e^{ (-\frac{\varepsilon(1-\lambda)\sqrt{\xi^{2}_1+(\xi_2-\xi_3)^2}}{2}-\frac{\varepsilon(1-\lambda)^2(\xi_2+\xi_3)^2}{4})} }{\int d^3\xi e^{-\varepsilon(\sqrt{\xi_1^2+\xi_2^2}-\xi_3^2)}} \\
	&=\frac{4}{(1-\lambda)^3},
\end{align*}
Using Eqn.(\ref{n3noisyprojrhowigner2}) the negativity can be calculated as
\begin{align}\label{n4noisyprojneg1}
	\mathscr{W}_{neg}^{proj++-}(\rho,\vec{A})=\frac{(1-\lambda)^3}{4}\mathscr{W}_{neg}^{mub}(\rho,\vec{A}).
\end{align}
Here also we see that as the noise parameter increases the negativity of the Wigner function decreases. If instead of Eqn.(\ref{n3noisyprojrhowigner2}) we use Eqns.(\ref{wigproj+-+},\ref{wigproj+--},\ref{wigproj-+-}) in the expression for the negativity of Wigner distribution in Eqn.(\ref{n3negcalproj}) we will retrieve Eqn.(\ref{n4noisyprojneg1}) exactly
\subsection{Four operators (n=4)}
\subsubsection{Calculating the Wigner function}
Here we will consider all the four noisy eigenprojections of $\sigma_x$ and $\sigma_y$ as our operators and calculate the Wigner distribution function. Our operator set is
\begin{itemize}
    \item $\{(\frac{\lambda}{2}I+(1-\lambda)\vert + \rangle_1 \langle + \vert),\frac{\lambda}{2}I+(1-\lambda)\vert - \rangle_1 \langle - \vert),(\frac{\lambda}{2}I+(1-\lambda)\vert + \rangle_2 \langle + \vert),(\frac{\lambda}{2}I+(1-\lambda)\vert - \rangle_2 \langle - \vert)\} $
\end{itemize}
Tracing the same steps as the previous calculations,  for a qubit density matrix $\rho=(\mathbb{I}+\sum_kr_k\sigma_k)/2$, where $\vec{r}=\{r_1,r_2\}$ is the Bloch vector, the Wigner function can be written as
\begin{align}\label{n4projidentorhoproj}
	\mathscr{W}_{\rho}^{proj+-+-}(\vec{a})=\frac{1}{2}(1+\vec{r} \cdot \vec{a}_{12}')\mathscr{W}_{\mathbb{I}}^{proj+-+-}(\vec{a}),
\end{align}
where $a'_{12}=(a'_1,a'_2), \vec{a}=\{a_1,a_2,a_3,a_4\}$ and
\begin{align}
    a'_1=\frac{a_1-a_2}{(1-\lambda)},\nonumber\\
    a'_2=\frac{a_3-a_4}{(1-\lambda)}.
\end{align}
The Fourier transform of the Wigner function, $\hat{\mathscr{W}_{\mathbb{I}}^{proj+-+-}(a_1,a_2,a_3,a_4)}$ comes out to be $2e^{i\frac{(\xi_{1}+\xi_{2}+\xi_{3}+\xi_{4})}{2}}\cos(\frac{(\lambda-1}{2}|\xi"_{12}|)$ where $|\xi"_{12}|=\sqrt{(\xi_1-\xi_2)^2+(\xi_3-\xi_4)^2}$. By convoluting it with a regularizing function of the form $e^{ (-\frac{\varepsilon(1-\lambda)|\xi"_{12|}}{2}-\frac{\varepsilon(1-\lambda)^2(\xi_1+\xi_2)^2+(\xi_3+\xi_4)^2}{4})}$. Using the transformation
\begin{align}
    \frac{(1-\lambda)}{2}(\xi_1-\xi_2)=\xi"_1,\nonumber\\
    \frac{(1-\lambda)}{2}(\xi_3-\xi_4)=\xi"_2,\nonumber\\
    \frac{(1-\lambda)}{2}(\xi_1+\xi_2)=\xi"_3,\nonumber\\
    \frac{(1-\lambda)}{2}(\xi_3+\xi_4)=\xi"_4.
\end{align}
The Wigner distribution can be calculated as
\begin{align}\label{n4noisyprojidenwigner}
	 \mathscr{W}_{\mathbb{I}}^{proj++-}(&\vec{a})=\frac{1}{(2 \pi)^{4}} \int d \xi e^{-i \xi a} 2e^{i\frac{(\xi_{1}+\xi_{2}+\xi_{3}+\xi_{4})}{2}}\nonumber \\
	 &\cos(\frac{(\lambda-1}{2}|\xi"_{12}|)
	 e^{ (-\frac{\varepsilon(1-\lambda)|\xi"_{12}|}{2}-\frac{\varepsilon(1-\lambda)^2(\xi_1+\xi_2)^2+(\xi_3+\xi_4)^2}{4})} \nonumber \\&=\frac{-4}{\pi (1-\lambda)^4}\delta(a'_3)\delta(a'_4)\frac{1}{(1-|a'_{12}|^2)^\frac{3}{2}}.
\end{align}
where 
\begin{align}
    a'_3=\frac{a_1+a_2-1}{(1-\lambda)}\nonumber\\
    a'_4=\frac{a_3+a_4-1}{(1-\lambda)}
\end{align}
Thus total Wigner function can be written as
\begin{equation}\label{n4noisyprojrhowigner}
    \mathscr{W}_{\rho}^{proj+-+-}(\vec{a})=\frac{1}{2}(1+\vec{r} \cdot \vec{a}_{12}')\frac{-4}{\pi (1-\lambda)^4}\delta(a'_3)\delta(a'_4)\frac{1}{(1-|a'_{12}|^2)^\frac{3}{2}}.
\end{equation}
\subsubsection{Calculating the negativity}
\begin{figure}[h!]
    \centering
    \includegraphics[scale=0.5]{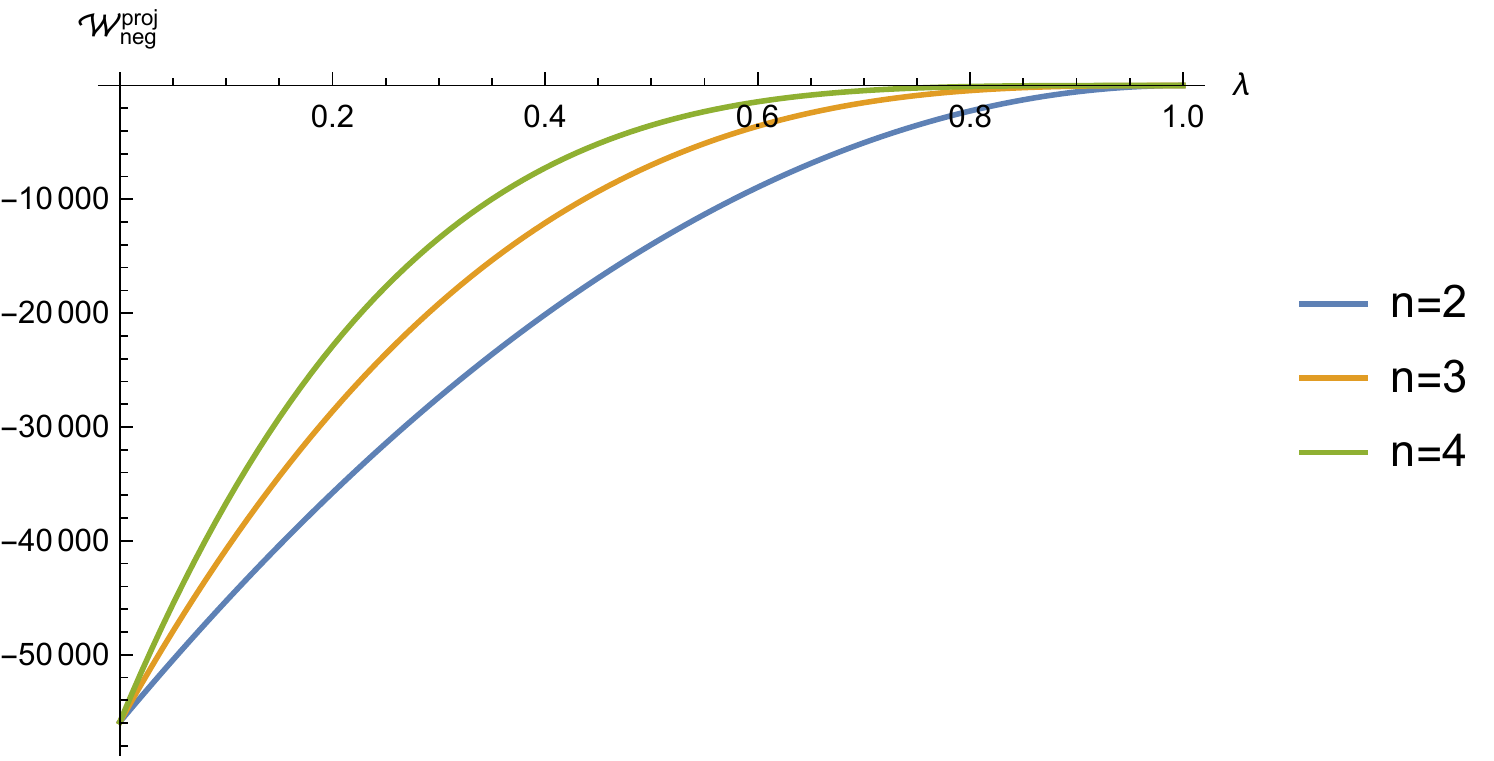}
    \caption{Variation of negativity of Wigner distribution function with noise $\lambda$ for n=2, n=3 and n=4 cases under scaled Gaussian regularization}
    \label{fig:comp}
\end{figure}

The negative volume of the Wigner distribution can be defined as
\begin{align}\label{n4negcalproj}
	\mathscr{W}_{neg}(\rho,\vec{A})= \int\frac{1}{2}(\mathscr{W}_{\rho}(\vec{a})-|\mathscr{W}_{\rho}(\vec{a})|)d^4a.
\end{align}
In accordance with the previous normalization requirement after dividing the Wigner distribution in Eqn.(\ref{n4noisyprojrhowigner}) with the factor
\begin{align*}
	\frac{I_{G'}}{I_{G}}&=\frac{\int d^4\xi e^{ (-\frac{\varepsilon(1-\lambda)\sqrt{\xi^{2}_1+(\xi_2-\xi_3)^2}}{2}-\frac{\varepsilon(1-\lambda)^2((\xi_1+\xi_2)^2+(\xi_3+\xi_4)^2)}{4})} }{\int d^4\xi e^{-\varepsilon(\sqrt{\xi_1^2+\xi_2^2}-\xi_3^2-\xi_4^2)}} \\
	&=\frac{4}{(1-\lambda)^4},
\end{align*}
The negative volume of the Wigner distribution can be calculated as
\begin{align}\label{n4noisyprojneg}
	\mathscr{W}_{neg}^{proj+-+-}(\rho,\vec{A})=\frac{(1-\lambda)^4}{4}\mathscr{W}_{neg}^{mub}(\rho,\vec{A}).
\end{align}
Thus we observe that by increasing the value of noise parameter $\lambda$ we can decrease the negative value of the Wigner distribution. Numerically integrating Eqns.(\ref{n2noisyprojneg1}),(\ref{n4noisyprojneg1}) and (\ref{n4noisyprojneg}) we get the plots in Fig.(\ref{fig:comp}). We see that for all three cases: n=2,3,4, the negativity of the Wigner function decreases with an increase in the noise. We also observe that for a non-zero value of the noise parameter, the negative volume is highest for the n=2 case and lowest for the n=4 case. This observation again supports the fact that with the addition of the noise among operators their incompatibility decreases and the Wigner function is capturing it through its negative volume.

Now, by using a little algebra, we can show that
\begin{align}
    \frac{\lambda}{2}I+(1-\lambda)\vert \pm \rangle_1 \langle \pm \vert)=\frac{I}{2}\pm\frac{(1-\lambda_k)}{2}\sigma_1\\
    \frac{\lambda}{2}I+(1-\lambda)\vert \pm \rangle_2 \langle \pm \vert)=\frac{I}{2}\pm\frac{(1-\lambda_k)}{2}\sigma_2\\
    \frac{\lambda}{2}I+(1-\lambda)\vert \pm \rangle_3 \langle \pm \vert)=\frac{I}{2}\pm\frac{(1-\lambda_k)}{2}\sigma_3
\end{align}
Hence, in principle, instead of analyzing the noisy eigenprojections of Pauli operators, we can just deal with the noisy Pauli operators themselves to see the trend of the negative volume of their corresponding Wigner distribution with the noise parameter. As we will see in the later sections, using noisy Pauli operators will be helpful in generalizing the treatment to the higher dimensions. Also dealing with Pauli operators is physically much more relevant because they can act as observables in a given system. For example, for electrons, they are proportional to the spin operators and for photon beams, they correspond to the measurement of Stokes' parameters. So for the remaining part of this paper, we will deal with noisy Pauli qubits and, in general, qudit operators.
\section{Wigner distribution function for Noisy Qubit operators}\label{Sec5}
 In this section, we add some noise to the both $n=3$ and $n=2$ Pauli operators and calculate the Wigner distribution function. We will then calculate the negative volume of the corresponding Wigner distribution function and analyze how it varies with the noise parameter in comparison to the noiseless case. First, we will study the n=3 case as it is easier to handle analytically as compared to the n=2 case. The Bloch vector is of the form $\vec{r}=\{r_1,r_2,r_3\}$. n=2 can be treated as a special case when $r_3=0$.

\subsection{Three operators (n=3)}
\subsubsection{Noisy Pauli Operators (Scaled Gaussian Regularization)}\label{asymm_noise}
Now we will consider the qubit operators to be, $A_k=(1-\lambda_k)\sigma_k+\lambda_k\mathbb{I}$ where $k=1,2,3$. As Eqn. (\ref{derivativeofWhat}) we have
\begin{align}\label{n3noisyidentorho}
	\mathscr{W}_{\rho}^{noisy}(\vec{a})=\frac{1}{2}(1+\vec{r} \cdot \vec{a}')\mathscr{W}_{\mathbb{I}}^{noisy}(\vec{a}),
\end{align}
where $a_k'=\frac{a_k-\lambda_k}{1-\lambda_k}$. Also, the Fourier transform of the Wigner distribution function changes to
\begin{align*}
\widehat{\mathscr{W}}_{\mathbb{I}}^{noisy}(\vec{\xi})=e^{i (\xi_{1} \lambda_1 +\xi_{2} \lambda_2+\xi_{3} \lambda_3)}2\cos(|\xi'|),
\end{align*}
where $|\xi'|=\sqrt{(1-\lambda_1)^{2}\xi^{2}_{1}+(1-\lambda_2)^{2}\xi^{2}_{2}+(1-\lambda_3)^{2}\xi^{2}_{3}}$. Now, doing the integration in Eq.(\ref{wignerdist}) is not so straight forward if we use the same Gaussian regularization function $e^{-\varepsilon\xi^2}$, because of lack of spherical symmetry in $\widehat{\mathscr{W}}_{\mathbb{I}}^{noisy}(\vec{\xi})$. Therefore we chose our regularization function to be a scaled Gaussian $G'=e^{-\varepsilon ((1-\lambda_1)^{2}\xi^{2}_{1}+(1-\lambda_2)^{2}\xi^{2}_{2}+(1-\lambda_3)^{2}\xi^{2}_{3})}$ then the calculations follows exactly the same way as for the Noiseless Pauli operators in \cite{schwonnek2020wigner} by changing the integration variables as $\xi_k\rightarrow\xi_k'=(1-\lambda_k)\xi_k$. The integration goes as follows 
\begin{align}\label{n3noisyidenwigner}
	 \mathscr{W}_{\mathbb{I}}^{noisy}(\vec{a})&=\frac{1}{(2 \pi)^{3}} \int d \xi e^{-i \xi a} e^{-i (\xi_{1} \lambda_1 +\xi_{2} \lambda_2+\xi_{3} \lambda_3)}2\cos(|\xi'|)e^{-\varepsilon\xi'^2} \nonumber \\&=\frac{-1}{\pi |a'|(1-\lambda_1)(1-\lambda_2)(1-\lambda_3)} \frac{d}{d |a'|} \frac{e^{-\frac{(|a'|-1)^{2}}{4 \varepsilon}}}{\sqrt{4 \pi \varepsilon}}  \nonumber \\&=\frac{-\delta^{\prime}(|a'|-1)}{2 \pi |a'|(1-\lambda_1)(1-\lambda_2)(1-\lambda_3)}.
\end{align}
Therefore, the Wigner distribution function for an arbitrary qubit state is given as
\begin{align}\label{n3noisyrhowigner}
	\mathscr{W}_{\rho}^{noisy}(\vec{a})=-\frac{(1+\vec{r} \cdot \vec{a}')}{4 \pi|a'|(1-\lambda_1)(1-\lambda_2)(1-\lambda_3)} \delta^{\prime}(|a'|-1)
\end{align}
It should be noted that the derivative of the $\delta-$ function in the above equation is with respect to $|a'|$ variable unlike in \cite{schwonnek2020wigner}, where the derivative is with respect to $|a|$.

\subsubsection{Noisy Pauli Operators with equal noise (Gaussian Regularization)}\label{symm_noise}
In the previous section, if the noise added to each of the Pauli operators is equal we can calculate the Wigner distribution function $\mathscr{W}^{noisy}_{\rho}(\mathbb{I})$ with the Gaussian regularization function also. In this case, the qubit operators are, $A_k=(1-\lambda)\sigma_k+\lambda\mathbb{I}$. The Wigner distribution function for an arbitrary density matrix can be written in terms of the Wigner distribution function  of $\mathbb{I}$ as 
\begin{align}\label{n3symmidentorho}
	\mathscr{W}_{\rho}^{symm}(\vec{a})=\frac{1}{2}(1+\vec{r} \cdot \vec{a}')\mathscr{W}_{\mathbb{I}}(\vec{a}),
\end{align}
where $a_k'=\frac{a_k-\lambda}{1-\lambda}$. The Fourier transform of the Wigner distribution function is given as 
\begin{align*}
	\widehat{\mathscr{W}}_{\mathbb{I}}^{symm}(\vec{\xi})=e^{i (\xi_{1} \lambda +\xi_{2} \lambda+\xi_{3} \lambda)}2\cos(|\xi'|),
\end{align*}
where $|\xi'|=(1-\lambda)\sqrt{\xi^{2}_{1}+\xi^{2}_{2}+\xi^{2}_{3}}$. The Wigner distribution function can now be calculated with the Gaussian regularization as 
\begin{align}\label{n3noisyidenwignersymm}
	\mathscr{W}_{\mathbb{I}}^{symm}(\vec{a})&=\frac{1}{(2 \pi)^{3}} \int d \xi e^{-i \xi a} e^{-i (\xi_{1} \lambda +\xi_{2} \lambda+\xi_{3} \lambda)}2\cos(|\xi'|)e^{-\varepsilon\xi^2} \nonumber \\
    & =\frac{-1}{\pi |a'|(1-\lambda)^2} \frac{d}{d |a'|} \frac{e^{-\frac{(|a'|-1)^{2}(1-\lambda)^2}{4 \varepsilon}}}{\sqrt{4 \pi \varepsilon}}  \nonumber \\
	&= \frac{-\delta^{\prime}(|a''|-(1-\lambda))}{2 \pi |a''|}.
\end{align}
where we have done the variable substitution $\xi\rightarrow\xi'$ and $a_k''=a_k-\lambda$.

Therefore, the Wigner distribution function for an arbitrary qubit state with equally noisy Pauli operators is given as
\begin{align}\label{n3noisyrhowignersymm}
	\mathscr{W}_{\rho}^{symm}(\vec{a})=-\frac{(1+\vec{r} \cdot \vec{a}')}{4 \pi|a''|} \delta^{\prime}(|a''|-(1-\lambda)).
\end{align}

We note that in the limit $\lambda_1=\lambda_2=\lambda_3=\lambda=0$ both Eqn.(\ref{n3noisyrhowigner}) and Eqn.(\ref{n3noisyrhowignersymm}) reduces to the expression for the Wigner function for n=3 qubit operator case in \cite{schwonnek2020wigner}.
\subsubsection{Arbitrary qubit operators}\label{section_arboperators}
Now, we will consider the following three-qubit operators which could represent a generic set of three-qubit Hermitian operators with eigenvalues $\pm1$
\begin{align}\label{arboperators}
    &A_1= \lambda_x\sigma_x+\lambda_y\sigma_y+\lambda_z\sigma_z, \nonumber \\
    &A_2= \gamma_x\sigma_x+\gamma_y\sigma_y+\gamma_z\sigma_z, \nonumber \\
    &A_3= \sigma_z,
\end{align}
where $\lambda_x^2+\lambda_y^2+\lambda_z^2=\gamma_x^2+\gamma_y^2+\gamma_z^2=1$, which ensures that the eigenvalues of $A_1$ and $A_2$ are $\pm1$. Note that because of unitary freedom, we can always choose one of the operators to be $\sigma_z$.

Next, we note that the Wigner distribution function of an arbitrary operator can be expanded as
\begin{align}\label{n3arbidentorho}
    \mathscr{W}_{\rho}^{arb}(\vec{a})=\frac{1}{2}(1+\vec{r} \cdot \vec{a}')\mathscr{W}_{\mathbb{I}}^{arb}(\vec{a}),
\end{align}
where the s parameters are given as 
\begin{align}\label{sparameters}
    &a'_1=\frac{-a_1\gamma_y+a_2\lambda_y+a_3(\gamma_y\lambda_z-\gamma_z\lambda_y)}{\gamma_x\lambda_y-\lambda_x\gamma_y},\nonumber\\
    &a'_2=\frac{a_1\gamma_x-a_2\lambda_x+a_3(\gamma_z\lambda_x-\gamma_x\lambda_z)}{\gamma_x\lambda_y-\lambda_x\gamma_y},\nonumber \\
    &a'_3=a_3.
\end{align}
Moreover, the Fourier transform of the Wigner distribution function takes the following form 
\begin{align}
\widehat{\mathscr{W}}_{\mathbb{I}}^{arb}(\vec{\xi})=\tr e^{i\xi_i A_i}=\tr e^{i(\eta_x \sigma_x+\eta_y
\sigma_y+\eta_z \sigma_z)}=\cos(|\eta|),
\end{align}
where the parameters $\vec{\eta}$ are given as $
    \eta_x=\xi_1\lambda_x+\xi_2\gamma_x,
    \eta_y=\xi_1\lambda_y+\xi_2\gamma_y,
    \eta_z=\xi_1\lambda_z+\xi_2\gamma_z+\xi_3.$ As before, our $\widehat{\mathscr{W}}_{\mathbb{I}}^{arb}(\vec{\xi})$ is not a symmetric function of $\vec{\xi}$. Now, we pick our regularisation function to be Gaussian with respect to $\vec{\eta}$ parameters, i.e., $G''=e^{-\varepsilon \eta^2}$ with $\eta^2=n_x^2+n_y^2+n_z^2$. With this regularisation function, we calculate the Wigner function of identity as
    \begin{align}\label{n3arbidenwigner}
        \mathscr{W}_{\mathbb{I}}^{arb}(\vec{a})&=\frac{1}{(2\pi)^3}\int d \xi e^{-i \vec{\xi} \cdot \vec{a}} 2\cos(|\eta|)e^{-\varepsilon\eta^2} \nonumber \\&=\frac{-\delta^{\prime}(|a|-1)}{2 \pi |a||(\gamma_y\lambda_x-\gamma_x\lambda_y)|}.
    \end{align}
Therefore the Wigner distribution function for an arbitrary qubit state for arbitrary qubit operators is given as    
\begin{align}\label{n3arbrhowigner}
	\mathscr{W}_{\rho}^{noisy}(\vec{a})=-\frac{(1+\vec{r} \cdot \vec{a}')}{4 \pi|a||(\gamma_y\lambda_x-\gamma_x\lambda_y)|} \delta^{\prime}(|a'|-1).
\end{align}

\subsubsection{Comparing the negativity}\label{negcompqubit}
We will calculate the negative volume of the Wigner function for noisy Pauli operators in the same way as in Sec.(\ref{Sec4}). Both scaled Gaussian regularisation and Gaussian regularisation approaches are treated one by one in order to compare the findings.

\textit{Noisy operators with arbitrary noise}: For the case of Noisy Pauli operators with arbitrary noise (\ref{asymm_noise}) we note that we had used the scaled Gaussian function for regularizing the Wigner distribution function. In order to compare its negativity, we take an average weight by dividing $\mathscr{W}_{\rho}^{noisy}(a)$ by the following quantity
\begin{align*}
	\frac{I_{G'}}{I_{G}}&=\frac{\int d^3\xi e^{-\varepsilon ((1-\lambda_1)^{2}\xi^{2}_{1}+(1-\lambda_2)^{2}\xi^{2}_{2}+(1-\lambda_3)^{2}\xi^{2}_{3})} }{\int d^3\xi e^{-\varepsilon(\xi_1^2+\xi_2^2+\xi_3^2)}} \\
	&=\frac{1}{(1-\lambda_1)(1-\lambda_2)(1-\lambda_3)},
\end{align*}
where $I_G'$ is the integral of the scaled Gaussian and $I_G$ is the integral of the Gaussian function over the Fourier space.
After this operation the expression for the negative volume of the Wigner function is evaluated as
\begin{align}\label{n3noisyneg}
	\mathscr{W}_{neg}^{noisy}(\rho,\vec{A})=(1-\lambda_1)(1-\lambda_2)(1-\lambda_3)\mathscr{W}_{neg}^{mub}(\rho,\vec{A}).
\end{align}
Thus, the negative volume for noisy operators is lesser than the negative volume for MUB operators, and with an increase in noise, it decreases further. 

\begin{figure}
    \centering
    \includegraphics[scale=0.5]{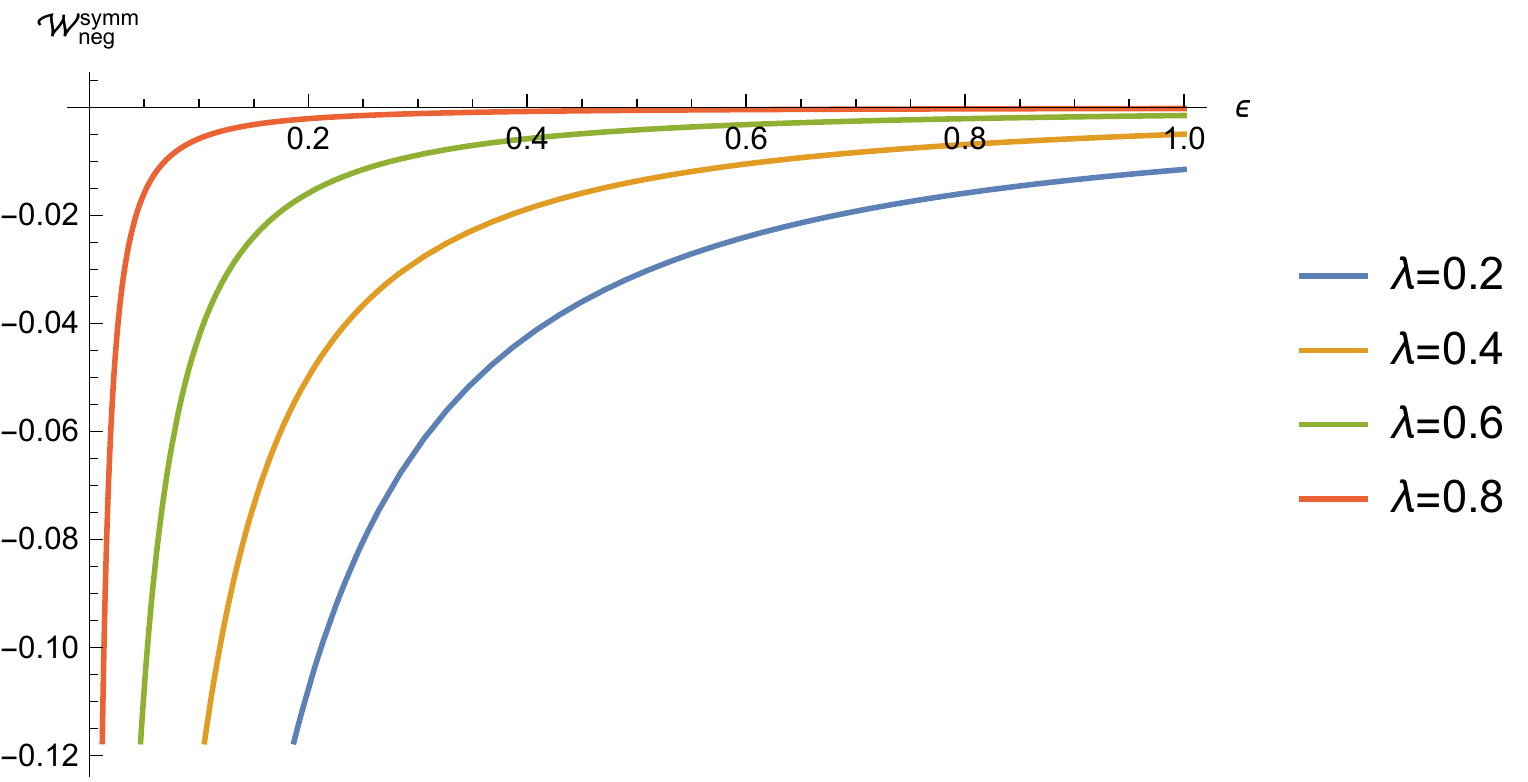}
    \caption{Variation of negativity of Wigner distribution function with regularising parameter $\varepsilon$ for different amounts of noise $\lambda$ under Gaussian regularization}
    \label{fig:3}
\end{figure}

\begin{figure}
    \centering

    \includegraphics[scale=0.5]{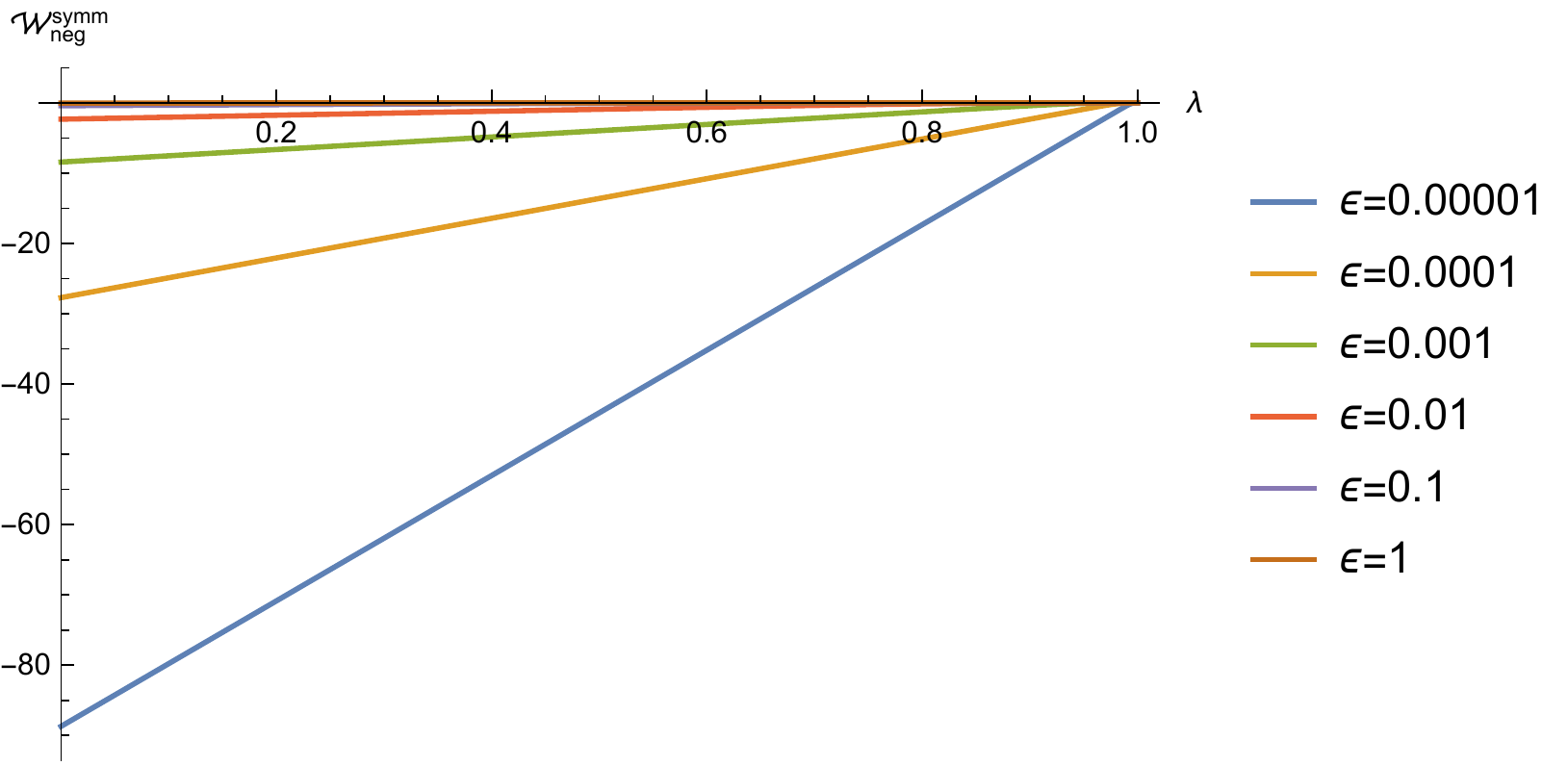}
    \caption{Variation of negativity of Wigner distribution function with noise $\lambda$ for different values of regularising parameter $\varepsilon$ under Gaussian regularization}
    \label{fig:4}
\end{figure}

\textit{Noisy Pauli operators with equal noise}:
Now we consider the case of noisy operators with equal noise (\ref{symm_noise}). In this case, the regularization function is the same as the one used for calculating the Wigner distribution of MUBs\cite{schwonnek2020wigner}. We calculate the negative volume of $\mathscr{W}_{\mathbb{\rho}}^{mub}(\vec{a})$ and $\mathscr{W}_{\mathbb{\rho}}^{symm}(\vec{a})$ by using the Gaussian approximation of delta function which we encountered while deriving Eqn.(\ref{n3noisyidenwignersymm}). Eqn.(\ref{n3noisyneg}) than reduces to
\begin{equation}\label{negnoise}
        \mathscr{W}_{neg}^{symm}(\rho,\vec{A})=\frac{1}{2}\left(-1+\frac{\lambda-1}{\sqrt{\pi \varepsilon}}+Erfc \left(\frac{\lambda-1}{2\sqrt{\varepsilon}}\right) \right).
    \end{equation}
We have plotted the results in Fig.(\ref{fig:3}) and Fig.(\ref{fig:4}) respectively. It can be seen that with increasing value of noise $\lambda$ the difference between the negative volume of $\mathscr{W}_{\rho}^{symm}(a)$ and $	\mathscr{W}_{\rho}^{mub}(a)$ keeps increasing as $\varepsilon\rightarrow0$. This indicates that the  negative volume for the maximally compatible case is the least. Similarly, for fixed values of $\varepsilon$ the negative volume keeps decreasing as  $\lambda\rightarrow1$. We can also use Eqn. (\ref{n3noisyneg}), where we used a scaled Gaussian regularisation function, to analyze how negative volume varies with noise parameter $\lambda$ and regularization parameter $\varepsilon$ by setting $\lambda_1=\lambda_2=\lambda_3=\lambda$. We get the following expression:
 \begin{equation}
        \mathscr{W}_{neg}^{noisy}(\rho,\vec{A})=\frac{(\lambda-1)^{3}}{2}\left(-1+\frac{1}{\sqrt{\pi \varepsilon}}+Erfc \left(\frac{1}{2\sqrt{\varepsilon}}\right) \right).
    \end{equation}
By plotting this in Fig.(\ref{fig:3}) and Fig.(\ref{fig:4}) we observe that we get similar trends as we got in the previous cases. This is an indication of the claim that using the scaled Gaussian regularisation to calculate the convoluted Wigner function and then normalizing it by a certain factor to compare its negative volume with the MUB case is the justified thing to do. Another thing that supports this claim is that the regularization used for obtaining Eqn.(\ref{n3noisyneg}) is just like taking a waited average of the Wigner function over the phase space if the Gaussian regularization is considered as the simple average. We know that the physical equation among two samples remains the same regardless of what kind of average of data sets of those samples we are considering. \textit{i.e} all the averages like simple average, weighted average, root mean square average. etc. will give us the same physical results only differing in relative numerical values. We again stress this point that under this procedure, we can't quantify the degree of incompatibility by the obtained numerical value of the negativity of the Wigner function. What we can extract is the relative trend of the negativity with the incompatibility. Thus we have a qualitative indication that the Wigner function indeed captures the incompatibility. Consequently in the remainder of this work, we will do the analysis by using properly normalized scaled regularisation.   

\textit{Set of arbitrary operators}: Here again, we had to use a scaled Gaussian distribution function to compute the Wigner distribution function. So we divide $\mathscr{W}_{\rho}^{arb}(a)$ by the following quantity 
\begin{align*}
    \frac{I_{G''}}{I_G}&=\frac{\int d^3\xi e^{-\varepsilon(\eta_1^2+\eta_2^2+\eta_3^2)}}{\int d^3\xi e^{-\varepsilon(\xi_1^2+\xi_2^2+\xi_3^2)}}\\
    &=\frac{1}{|(\gamma_y\lambda_x-\gamma_x\lambda_y)|},
\end{align*}
where $I_{G''}$ is the integral of the scaled Gaussian used to compute the Wigner distribution function for a set of arbitrary operators.  After this division, the negativity for the Wigner function can be calculated as follows
\begin{align}\label{n3arbrnoise}
	\mathscr{W}_{neg}^{arb}(\rho,\vec{A})=|(\gamma_y\lambda_x-\gamma_x\lambda_y)|\mathscr{W}_{neg}^{mub}(\rho,\vec{A}), 
\end{align}
where $|(\gamma_y\lambda_x-\gamma_x\lambda_y)|\leq 1$. Thus we find a decrease in the total negativity when we chose an arbitrary set of operators instead of mutually unbiased operators. 

It can further be shown that this negativity decreases as the operators come close to each other, using a geometric approach. This further supports the intuition that negativity is a proper indicator of incompatibility, even for three observables. Suppose the eigenvectors of the observables in \eqref{arboperators}, are along the directions $A_1=\{\theta_1, \phi_1\}$, $A_2=\{\theta_2,\phi_2\}$ and $A_3=\{0,0\}$ in the Bloch sphere and their eigenvalues are $\pm 1$. 
So that the coefficients can now be written as 
\begin{align*}
\lambda_x=\sin(\theta_1)\cos(\phi_1),\quad\lambda_y=\sin(\theta_1)\sin(\phi_1) \quad \text{and} \quad \lambda_z=\cos(\theta_1) \\
\gamma_x=\sin(\theta_2)\cos(\phi_2),\quad\gamma_y=\sin(\theta_2)\sin(\phi_2) \quad \text{and} \quad \gamma_z=\cos(\theta_2),
\end{align*}

Substituting the above relations in \eqref{n3arbrnoise}, we get that 
\begin{align}\label{n3arbrnoise_geom}
	\mathscr{W}_{neg}^{arb}(\rho,\vec{A})=|(\sin \theta_1\sin\theta_2\sin(\phi_1-\phi_2)|\mathscr{W}_{neg}^{mub}(\rho,\vec{A}), 
\end{align}

where the factor in the front becomes smaller as the lines containing the eigenvectors of the three operators come closer to each other, i.e.,  the solid angle between them becomes smaller. Intuitively, we know that if the lines 
 in the Bloch sphere containing the eigenvectors come close to each other, the observables become more and more compatible.

\begin{figure}
    \centering
    \includegraphics[scale=0.5]{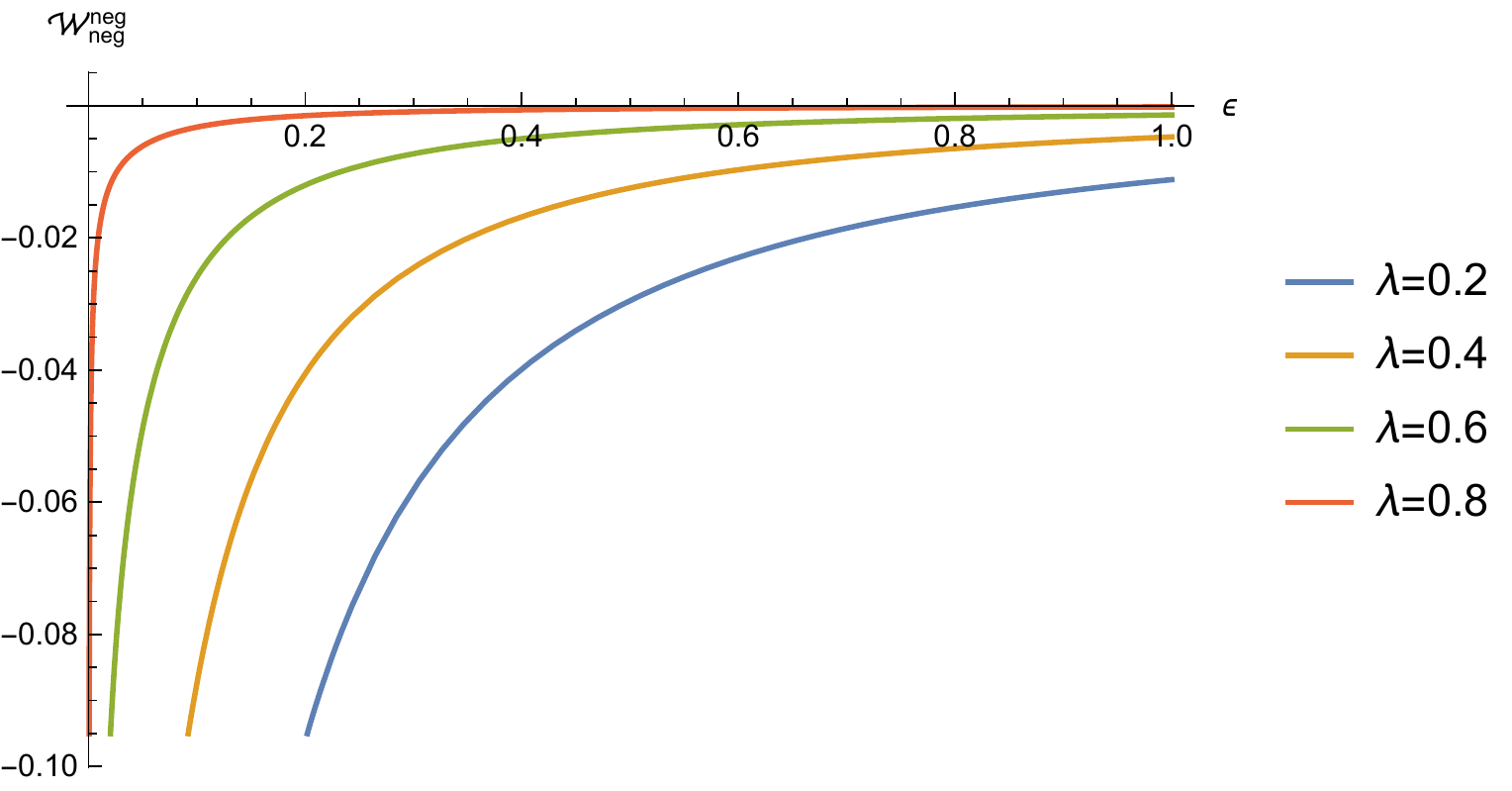}
    \caption{Variation of negativity of Wigner distribution function with regularising parameter $\varepsilon$ for different amounts of noise $\lambda$ under scaled Gaussian regularization}
    \label{fig:1}
\end{figure}
\begin{figure}
    \centering
    \includegraphics[scale=0.5]{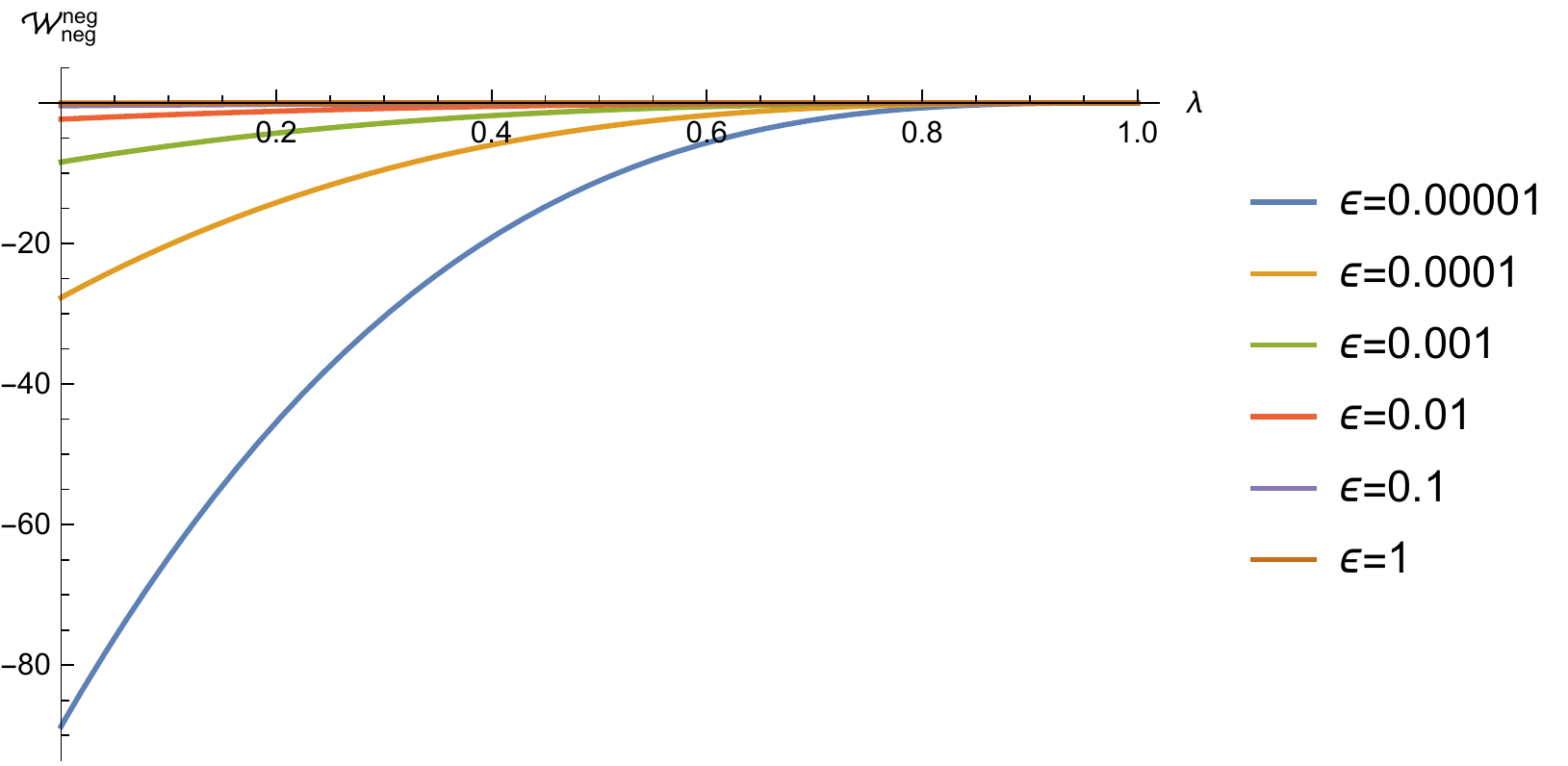}
    \caption{Variation of negativity of Wigner distribution function with noise $\lambda$ for different values of regularising parameter $\varepsilon$ under scaled Gaussian regularization}
    \label{fig:2}
\end{figure}
\subsection{Two operators (n=2)}
\subsubsection{Noisy Pauli Operators (Scaled Regularization)}\label{NosiyPauli}
In this case, the qubit operators are $A_k=(1-\lambda_k)\sigma_k+\lambda_k\mathbb{I},\ k=1,2$. Eqn.(\ref{n3noisyidentorho}) still holds with $r_3=0$. The Fourier transform of the Wigner function is now given by:
\begin{align*}
\widehat{\mathscr{W}}_{\mathbb{I}}^{noisy}(\vec{\xi})=e^{i (\xi_{1} \lambda_1 +\xi_{2} \lambda_2)}2\cos(|\xi'|),
\end{align*}
where $|\xi'|=\sqrt{(1-\lambda_1)^{2}\xi^{2}_{1}+(1-\lambda_2)^{2}\xi^{2}_{2}}$.  As we did in the n=3 case in order to exploit spherical symmetry to evaluate the convoluted Wigner function we choose a scaled regularisation function of the form $e^{-\varepsilon\sqrt{ ((1-\lambda_1)^{2}\xi^{2}_{1}+(1-\lambda_2)^{2}\xi^{2}_{2}}}$. The calculations then proceed in the same way as for the noiseless Pauli operators with the substitution  $\xi_k\rightarrow\xi_k'=(1-\lambda_k)\xi_k$. The integration follows as
\begin{align}
        \mathscr{W}_{\mathbb{I}}^{noisy}(\vec{a})&= \frac{1}{(2 \pi)^{2}} \int d\xi e^{-i\xi \cdot a}2 e^{-i (\xi_{1} \lambda_1 +\xi_{2} \lambda_2)}\cos{|\xi'|}e^{-\varepsilon |\xi'|}\nonumber\\
       &=\frac{-1}{(1-\lambda_1)(1-\lambda_2)\pi\left(1-|a'|^{2}\right)^{3 / 2}},
\end{align}
where $a'_k=\frac{a_k-\lambda_k}{1-\lambda_k}$. The calculation exactly follows the n=2 case in \cite{schwonnek2020wigner}. Using Eqn.(\ref{n3noisyidentorho}) with $r_3=0$ the total Wigner function is given as:
\begin{equation}\label{wignoisy2}
\mathscr{W}_{\rho}^{noisy}(\vec{a})=-\frac{(1+r \cdot a')}{2\pi(1-\lambda_1)(1-\lambda_2)}\frac{1}{\left(1-|a'|^{2}\right)^{3 / 2}}.
\end{equation}
Now we will observe how the calculations get much simpler if the noise added to each observable is equal.
\subsubsection{Noisy Pauli Operators with equal noise(Unscaled Regularization)}\label{Symmnoisen2}
As we have seen earlier if we add equal noise to each observable we can use unscaled regularization. The operator set becomes $A_k=(1-\lambda)\sigma_k+\lambda\mathbb{I}$. The Fourier transform of Wigner function comes out to be $\widehat{\mathscr{W}}_{\mathbb{I}}^{noisy}(\xi)=e^{i (\xi_{1} \lambda +\xi_{2} \lambda)}2\cos(|\xi'|)$ where $|\xi'|=(1-\lambda)\sqrt{\xi^{2}_{1}+\xi^{2}_{2}}$. The Wigner distribution can be calculated as:
\begin{align}
        \mathscr{W}_{\mathbb{I}}^{symm}(a)&= \frac{1}{(2 \pi)^{2}} \int d\xi e^{-i\xi \cdot a}e^{-i (\xi_{1} \lambda +\xi_{2} \lambda)}2\cos(|\xi'|)e^{-\varepsilon |\xi|}\nonumber\\
       &=\frac{-(1-\lambda)}{\pi\left((1-\lambda)^{2}-|a''|^{2}\right)^{3 / 2}},
\end{align}
where $a''_k=a_k-\lambda$. The total Wigner function is then given by:
\begin{equation}\label{wigsymm2}
\mathscr{W}_{\rho}^{symm}(\vec{a})=-\frac{((1-\lambda)+\vec{r} \cdot \vec{a}'')}{2\pi}\frac{1}{\left((1-\lambda)^2-|a''|^{2}\right)^{3 / 2}}.
\end{equation}
In this case also in the limit $\lambda_1=\lambda_2=\lambda=0$ both Eqn.(\ref{wignoisy2}) and Eqn.(\ref{wigsymm2}) reduces to the expression for Wigner function for n=2 qubit operator case in \cite{schwonnek2020wigner}.
\subsubsection{Arbitrary qubit operators}
Next, we will consider the following qubit operators which could represent a generic set of two-qubit Hermitian operators with eigenvalue $\pm1$
\begin{align}\label{arboperatorsn2}
    &A_1= \sigma_x, \nonumber \\
    &A_2= \sqrt{(1-\beta^{2})}\sigma_x+\beta\sigma_y,
\end{align}
The Wigner function for any arbitrary state can be written in terms of Wigner state of identity as:
\begin{align}\label{n2arbidentorho}
    \mathscr{W}_{\rho}^{arb}(\vec{a})=\frac{1}{2}(1+\vec{r} \cdot \vec{a}')\mathscr{W}_{\mathbb{I}}^{arb}(\vec{a}),
\end{align}
where
\begin{align}\label{aparameters}
    &a'_1=a_1,\nonumber\\
    &a'_2=\frac{a_2-\sqrt{(1-\beta^{2})}a_1}{\beta}.
\end{align}
The Fourier transform of the Wigner distribution function takes the following form
\begin{align*}
\widehat{\mathscr{W}}_{\mathbb{I}}^{arb}(\vec{\xi})=2\cos(|\eta|),
\end{align*}
where $\eta_1=\xi_1+\xi_2\sqrt{(1-\beta^{2}},  \eta_2=\xi_2\beta$. As before, $\widehat{\mathscr{W}}_{\mathbb{I}}^{arb}(\xi)$ is not a spherically symmetric function. So, in order to get the convoluted Wigner function we choose our regularizing function to be $e^{-\varepsilon|\eta|}$. Here $|\eta|=\sqrt{\eta_1^2+\eta_2^2}$ Using this the Wigner function of identity comes out to be
\begin{align}
        \mathscr{W}_{\mathbb{I}}^{arb}(\vec{a})&= \frac{1}{(2 \pi)^{2}} \int d\xi e^{-i\xi \cdot a}2\cos{|\eta|}e^{-\varepsilon |\eta|}\nonumber\\
       &= \frac{1}{\beta (2 \pi)^{2}} \int d\eta e^{-i\eta \cdot a'}2\cos{|\eta|}e^{-\varepsilon |\eta|}\nonumber\\
       &\rightarrow\frac{-1}{\beta \pi\left(1-|a'|^{2}\right)^{3 / 2}},
\end{align}
Using Eqn.(\ref{n2arbidentorho}) the Wigner distribution for any arbitrary qubit operator is given as:
\begin{equation}\label{wigarbsymm2}
\mathscr{W}_{\rho}^{symm}(\vec{a})=-\frac{(1+\vec{r} \cdot \vec{a}')}{2\pi \beta}\frac{1}{\left(1-|a'|^{2}\right)^{3 / 2}},
\end{equation}
where $|a'|^{2}$ is given by Eqn.(\ref{aparameters}).
\subsubsection{Comparing the negativity}\label{n2cal}
For this case also the negative volume of the Wigner function can be calculated using Eqn.(\ref{negcalproj}). We will compare the change in negativity with noise parameters for all three cases discussed above for qubits one by one.

\textit{Noisy Pauli operators with asymmetric noise}: For noisy operators when white noise is added asymmetrically (\ref{NosiyPauli}) just like the n=3, the expression of Wigner function needs to be normalized to compare its negativity to MUB case. We take an average weight by dividing $\mathscr{W}_{\rho}^{noisy}(\vec{a})$ by the following quantity
\begin{align*}
	\frac{I_{G'}}{I_{G}}&=\frac{\int d^2\xi e^{-\varepsilon\sqrt{ (1-\lambda_1)^{2}\xi^{2}_{1}+(1-\lambda_2)^{2}\xi^{2}_{2}} }}{\int d^2\xi e^{-\varepsilon\sqrt{\xi_1^2+\xi_2^2}}} \\
	&=\frac{1}{(1-\lambda_1)(1-\lambda_2)},
\end{align*}
where $I_{G'}$ is the integral of the scaled Gaussian and $I_G$ is the integral of the Gaussian function over the Fourier space. Thus using Eqn.(\ref{wignoisy2}) the negativity of Wigner function is evaluated as
\begin{align}\label{noisyneg12}
	\mathscr{W}_{neg}^{noisy}(\rho,\vec{A})=(1-\lambda_1)(1-\lambda_2)\mathscr{W}_{neg}^{mub}(\rho,\vec{A}).
\end{align}
Thus from the above expression, we see that the negative volume of the Wigner function for noisy operators is lesser than the corresponding negative volume for MUB and further decreases with an increase in the noise.

\begin{figure}
    \centering
    \includegraphics[scale=0.5]{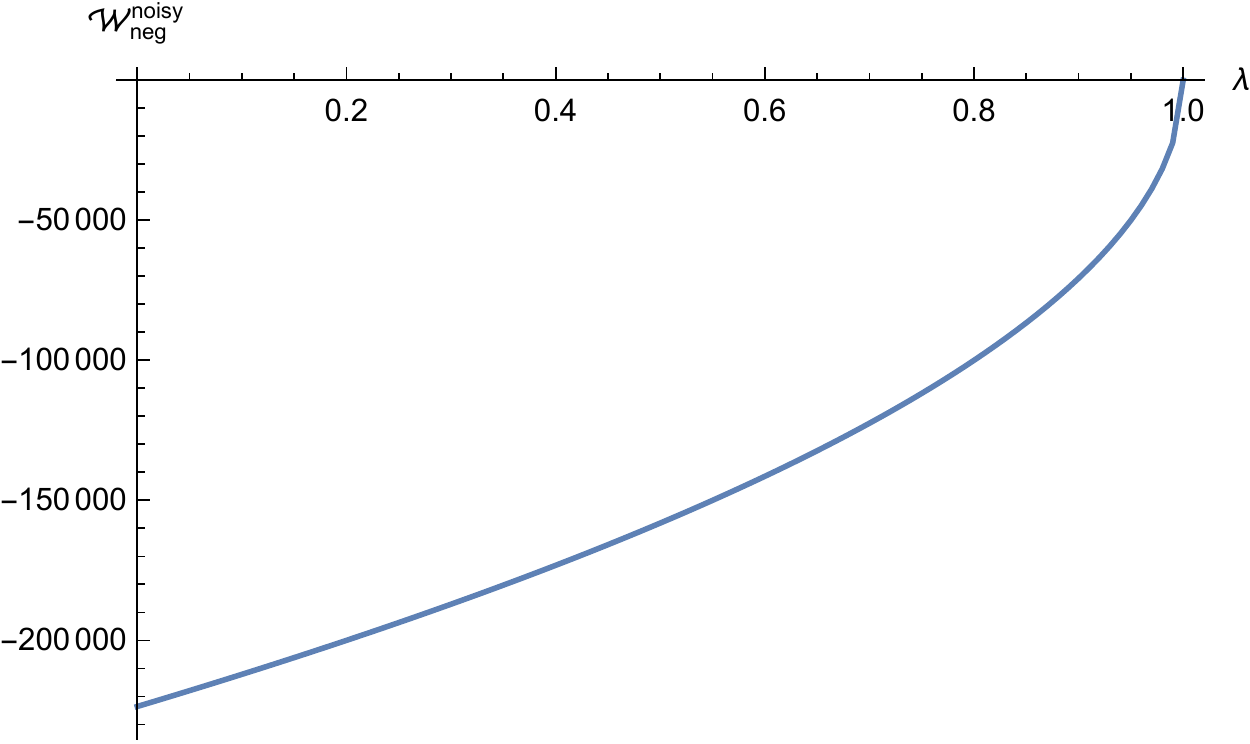}
    \caption{Variation of negativity of Wigner distribution function with noise $\lambda$   under Gaussian regularization}
    \label{fig:6}
\end{figure}

\begin{figure}
    \centering
    \includegraphics[scale=0.5]{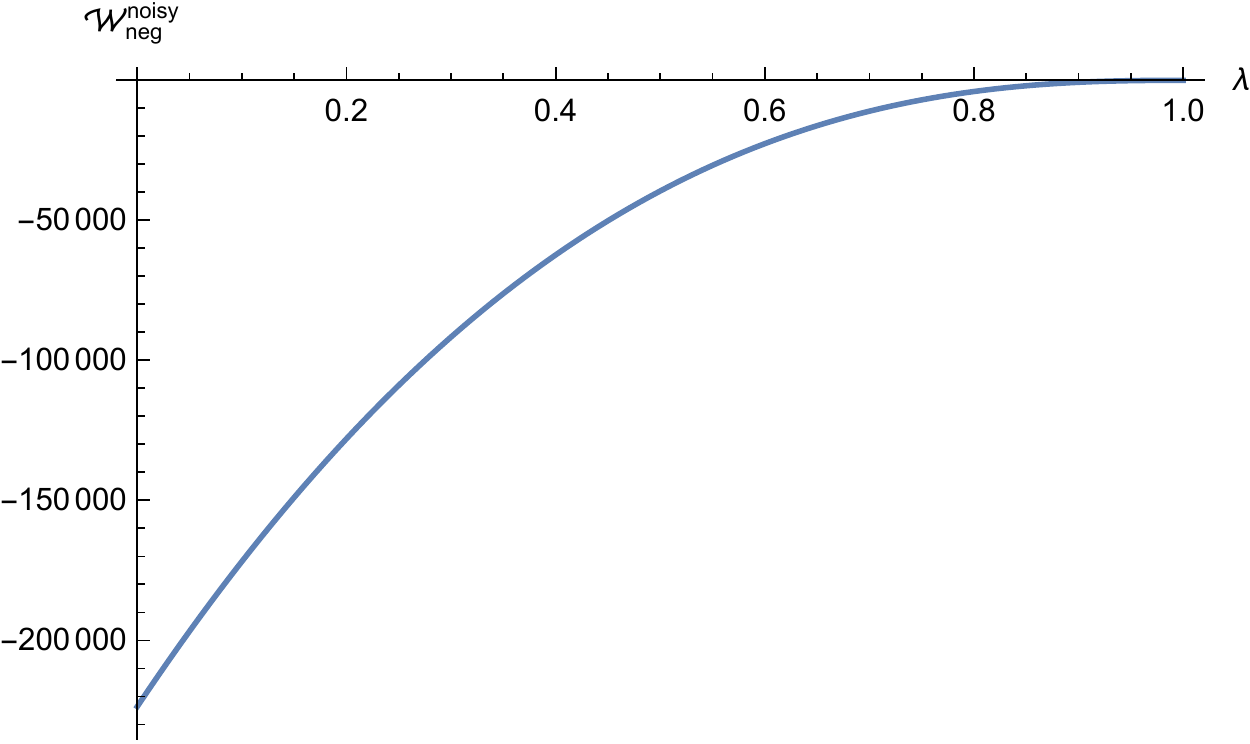}
    \caption{Variation of negativity of Wigner distribution function with noise $\lambda$  under scaled Gaussian regularization}
    \label{fig:7}
\end{figure}

\textit{Noisy Pauli operators with equal noise}: For noisy Pauli operators with equal noise we see from (\ref{Symmnoisen2}) that we can choose an unscaled regularization to get the expression for the Wigner function in Eqn.(\ref{wigsymm2}). We can use this expression and numerically integrate it to calculate the negative volume of the Wigner function against various amounts of noise. Fig.(\ref{fig:6}) is the result of this procedure. We see that the negative volume decreases with an increase in the noise parameter. As we know that with the addition of noise in the observables, the compatibility among them increases we observe that for the maximally compatibility case, vis-à-vis, the maximum value of the noise parameter, negative volume is the least. Similarly, instead of unscaled regularisation, we can consider scaled regularization. We can repeat the same procedure by using the expression of the Wigner function in Eqn.(\ref{wignoisy2}) with $\lambda_1=\lambda_2=\lambda$. We get Fig.(\ref{fig:7}) as a result of this. The trend of the negative volume against the noise parameter is similar to the one we got for the unscaled regularizing function. This is again an indication of the fact that we can compare the negative volume of the Wigner function for noisy operators to the negative volume for the MUB case by choosing a scaled regularizing function and then normalizing it properly.

\textit{Set of arbitrary operators}: Here again, we had to use a scaled regularizing function to compute the Wigner distribution function. Normalising expression in Eqn.(\ref{wigarbsymm2}) as:
\begin{align*}
	\frac{I_{G''}}{I_{G}}&=\frac{\int d^2\xi e^{-\varepsilon\sqrt{ (\xi_{1}+\xi_{2}(\sqrt{1-\beta^2}))^2+\xi^{2}_{2}\beta^2} }}{\int d^2\xi e^{-\varepsilon\sqrt{\xi_1^2+\xi_2^2}}} \\
	&=\frac{1}{|\beta|}.
\end{align*}
 Using that the negative volume comes out to be
\begin{align}\label{noisyneg1}
	\mathscr{W}_{neg}^{arb}(\rho,\vec{A})=|\beta|\mathscr{W}_{neg}^{mub}(\rho,\vec{A}),
\end{align}
where $\beta\leq1$. Hence we see that with an increase in beta, the negative volume increases and it becomes maximum when $\beta=1$ \textit{i.e.} which corresponds to the maximum incompatibility.

\begin{figure}
    \centering
    \includegraphics[scale=0.5]{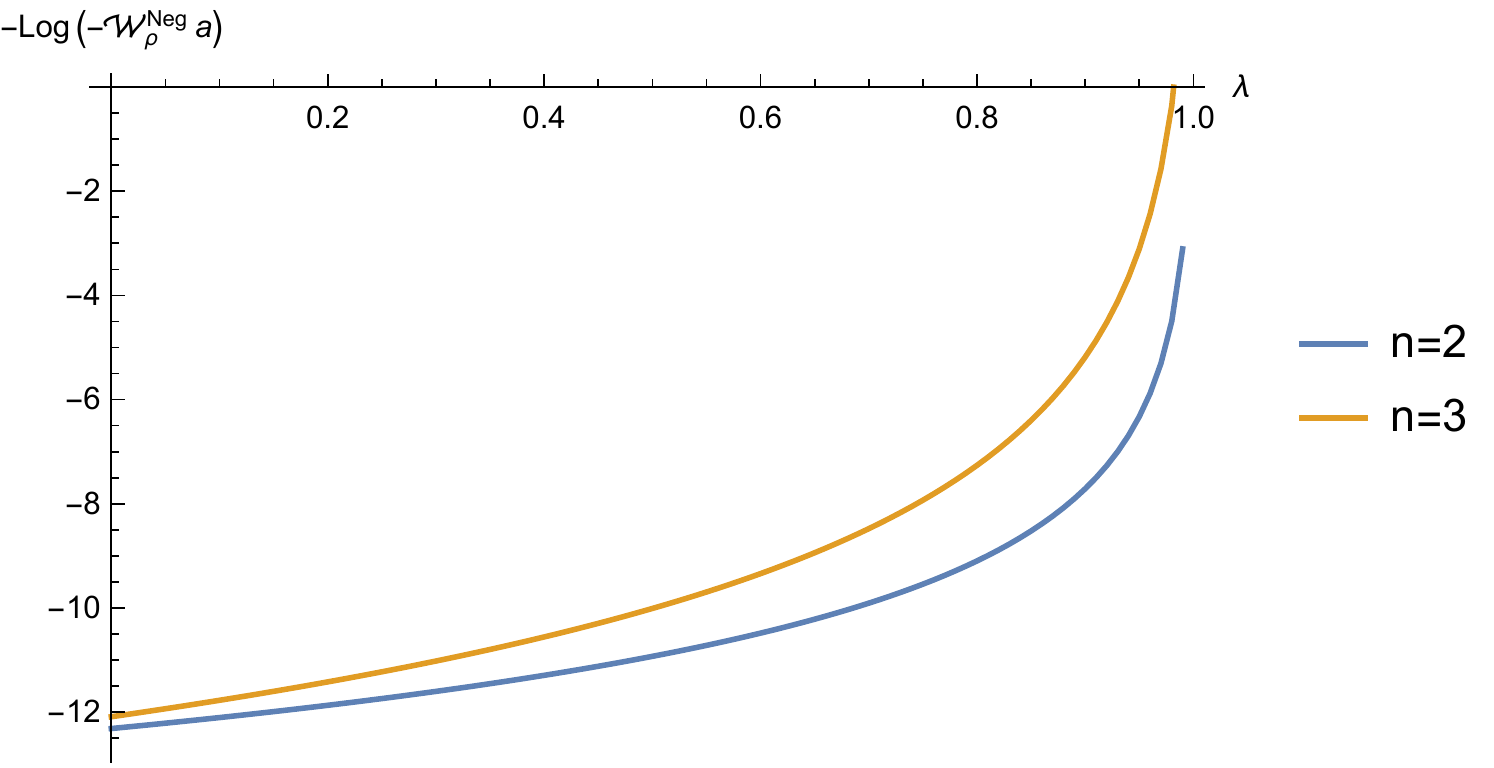}
    \caption{Comparison of the variation of negativity of the Wigner distribution with noise for n=2 and n=3 on Log scale. For $n=3$ case the value of regularising parameter is  $\varepsilon=0.0000001$ while for $n=2$ we have taken $\varepsilon\rightarrow 0$.}
    \label{fig:18}
\end{figure}

In Fig.(\ref{fig:18}) we compare the negative volume of the Wigner distribution for two and three operators for qubits although it is important to mention that this is not an absolute comparison as the regularisation function and the value of the regularisation parameter is different for both the cases.

\section{Wigner distribution function for the set of Qudit operators}\label{Sec7}
Till now we have analyzed 2-dimensional(Qubits) systems. In this section we will generalize this procedure to arbitrary but finite dimensions (Qudits), say N. We will calculate the Wigner function and negativity for various choices of Hermitian operators. We will also see how negativity changes with an increase in the dimension of the system. We start with three Hermitian qudit operators. \textit{i.e} we first consider n=3 case.
\subsection{Three operators (n=3)}
\subsubsection{Gell-Mann operators}
Gell-Mann operators are very good candidates for the generalization of our procedure to N dimensions. Their comparatively simpler forms in N dimensions allow us to calculate $\widehat{\mathscr{W}_{\mathbb{I}}}$ very efficiently. Also, its analytical form comes out to be similar to those we calculated in the previous sections for qubits. Thus the convoluted Wigner function and its negativity can be calculated and compared for any arbitrary dimension easily. Gell-Mann operators, as discussed earlier, are the generators of SU(N) for an N-dimensional qudit. Any density matrix can be decomposed using these operators as $ \rho=\frac{1}{N}(\mathbb{I}+\sum_{k}r_{k}\Lambda_{k}^N)$ where $r_k$ is now a N-dimensional Bloch vector and $\Lambda_k^N$ are N-dimensional Gell-Mann operators. The Wigner function for any arbitrary state can be written in terms of the Wigner function for the identity for a given set of operators as
\begin{equation}\label{Quditwigidentorho}
    \mathscr{W}_{\rho}(\vec{a})=\frac{1}{N}(1+\vec{r} \cdot \vec{a})\mathscr{W}_{\mathbb{I}}(\vec{a}),
\end{equation}
Let us consider our qudit operators to be $A_k=\Lambda_k^N, k=1,2,3$ such that
\begin{equation}
\begin{gathered}
\Lambda_{1}^N=\begin{pmatrix}
0 & 1 & 0 & \cdot & \cdot & \cdot & 0\\
1 & 0 & 0 & \cdot & \cdot & \cdot & 0\\
\cdot & \cdot & \cdot & \cdot &   &   & \cdot\\
\cdot & \cdot & \cdot &   & \cdot &   & \cdot\\
\cdot & \cdot & \cdot &   &   & \cdot & \cdot\\
0 & 0 & 0 & \cdot & \cdot & \cdot & 0
\end{pmatrix},
\qquad
\Lambda_{2}^N=\begin{pmatrix}
0 & -i & 0 & \cdot & \cdot & \cdot & 0\\
i & 0 & 0 & \cdot & \cdot & \cdot & 0\\
\cdot & \cdot & \cdot & \cdot &   &   & \cdot\\
\cdot & \cdot & \cdot &   & \cdot &   & \cdot\\
\cdot & \cdot & \cdot &   &   & \cdot & \cdot\\
0 & 0 & 0 & \cdot & \cdot & \cdot & 0
\end{pmatrix},\nonumber \\
\Lambda_{3}^N=\begin{pmatrix}
1 & 0 & 0 & \cdot & \cdot & \cdot & 0\\
0 & -1 & 0 & \cdot & \cdot & \cdot & 0\\
\cdot & \cdot & \cdot & \cdot &   &   & \cdot\\
\cdot & \cdot & \cdot &   & \cdot &   & \cdot\\
\cdot & \cdot & \cdot &   &   & \cdot & \cdot\\
0 & 0 & 0 & \cdot & \cdot & \cdot & 0
\end{pmatrix}.
\end{gathered}
\end{equation}
In that case Bloch vector $r_k=0$ for $k>3$. The expression for Fourier transform of the Wigner function for identity comes out to be $\widehat{\mathscr{W}_{\mathbb{I}}}(\vec{\xi})=(N-2)+2\cos{|\xi|}$ where $|\xi|=\sqrt{\xi_1^2+\xi_2^2+\xi_3^2}$. From this, we can calculate the Wigner function as
\begin{align}\label{Quditn3mubidenwigner}
	\mathscr{W}_{\mathbb{I}}(\vec{a})&=\frac{1}{(2 \pi)^{3}} \int d^3 \xi e^{-i \vec{\xi} \cdot \vec{a}} ((N-2)+2\cos(|\xi|))e^{-\varepsilon \xi^2}\nonumber\\
	&=\frac{-1}{2 \pi |a|} ((N-2)\delta^{\prime}(|a|)+\delta^{\prime}(|a|-1)).
\end{align}
Here we still are working in the limit $\varepsilon\rightarrow0$. Using Eqn.(\ref{Quditwigidentorho}) and Eqn.(\ref{Quditn3mubidenwigner}) we get
\begin{equation}\label{Quditn3mubrhowigner}
	\mathscr{W}_{\rho}(\vec{a})=\frac{-(1+\vec{r} \cdot \vec{a})}{2N\pi|a|} ((N-2)\delta^{\prime}(|a|)+\delta^{\prime}(|a|-1)).
\end{equation}
This expression reduces to the expression of the Wigner function for qubits when N=2 in \cite{schwonnek2020wigner}. 
\subsubsection{Noisy Gell-Mann Operators (Scaled Gaussian Regularization)}
Now we will inject noise in Gell-Mann operators as $A_k=(1-\lambda_k)\Lambda_k^N+\lambda_k\mathbb{I}$. For these operator Eqn.(\ref{Quditwigidentorho}) transforms to
\begin{equation}\label{Quditwigidentorhonoise}
    \mathscr{W}_{\rho}^{noisy}(\vec{a})=\frac{1}{N}(1+\vec{r} \cdot \vec{a}^{\prime})\mathscr{W}_{\mathbb{I}}^{noisy}(\vec{a}^{\prime}),
\end{equation}
where $a_{k}^{\prime}=\frac{a_{k}-\lambda_{k}}{1-\lambda_{k}}$. The Fourier transform of the Wigner function modifies to $\widehat{\mathscr{W}}_{\mathbb{I}}^{noisy}(\vec{\xi})=e^{i (\xi_{1} \lambda_1 +\xi_{2} \lambda_2+\xi_{3} \lambda_3)}((N-2)+2\cos(|\xi'|))$ with $|\xi'|=\sqrt{(1-\lambda_1)^{2}\xi^{2}_{1}+(1-\lambda_2)^{2}\xi^{2}_{2}+(1-\lambda_3)^{2}\xi^{2}_{3}}$. Here again, as in qubit cases, we choose our regularization function to be a scaled version of Gaussian regularization function $G'e^{-\varepsilon ((1-\lambda_1)^{2}\xi^{2}_{1}+(1-\lambda_2)^{2}\xi^{2}_{2}+(1-\lambda_3)^{2}\xi^{2}_{3})}$ to find the convoluted Wigner function. The spherical symmetry obtained by doing this allows the calculations to follow the same course as in noiseless Gell-Mann operators with the integration parameter being modified as $\xi_k\rightarrow\xi'_k=(1-\lambda_k)\xi_k$. The integration proceeds as
\begin{align}\label{Quditn3noisyidenwigner}
	 \mathscr{W}_{\mathbb{I}}^{noisy}(\vec{a})&=\frac{1}{(2 \pi)^{3}} \int d^3 \xi e^{-i \vec{\xi} \cdot \vec{a}} ((N-2)+2\cos(|\xi'|))e^{-\varepsilon\xi'^2} \nonumber \\&=\frac{-((N-2)\delta^{\prime}(|a'|)+\delta^{\prime}(|a'|-1))}{2 \pi |a'|(1-\lambda_1)(1-\lambda_2)(1-\lambda_3)}.
\end{align}
From this we calculate the Wigner function for an arbitrary state as
\begin{align}\label{Quditn3noisyrhowigner}
	\mathscr{W}_{\rho}^{noisy}(\vec{a}') =-\frac{(1+\vec{r} \cdot \vec{a}')((N-2)\delta^{\prime}(|a'|)+\delta^{\prime}(|a'|-1))}{2N\pi|a'|(1-\lambda_1)(1-\lambda_2)(1-\lambda_3)}.
\end{align}
Once again, the derivative of both the $\delta$ functions in the above equation is with respect to variable $|a'|$. We note that this expression for the Wigner function reduces to Eqn.(\ref{n3noisyrhowigner}) for N=2.
\subsubsection{Comparing the negativity}
For general Qudit cases, we can again proceed in the same way as we did for qubit cases. Using Eqn.(\ref{n3negcalproj}) and Eqn.(\ref{Quditn3mubrhowigner}) the negative volume for the noiseless case thus can be calculated as:
\begin{align}\label{mubnegqutrit}
	\mathscr{W}_{neg}(\rho,\vec{A})&= \int\frac{1}{2}\Big(\mathscr{W}_{\rho}(\vec{a})-|\mathscr{W}_{\rho}(\vec{a})|\Big)d^3a  \nonumber\\
	&= \int\frac{1}{2}\Bigg(\frac{(1+\vec{r} \cdot \vec{a})}{2N\pi|a|} ((N-2)\delta^{\prime}(|a|)+\delta^{\prime}(|a|-1)) \nonumber\\
	&-\Big|\frac{(1+\vec{r} \cdot \vec{a})}{2N\pi|a|} ((N-2)\delta^{\prime}(|a|)+\delta^{\prime}(|a|-1))\Big|\Bigg)d^3a.
\end{align}
Next will compare the noisy scenarios with this expression

\begin{figure}
    \centering
    \includegraphics[scale=0.7]{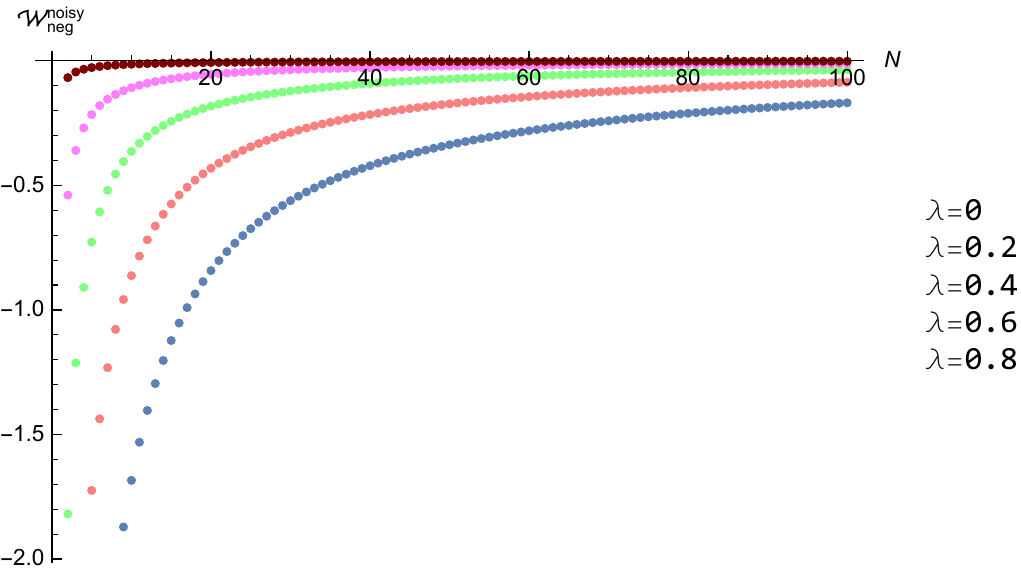}
    \caption{Variation of negativity of Wigner distribution function with the dimension of the operators N for different amounts of noise $\lambda$ under Scaled Gaussian regularization}
    \label{fig:12}
\end{figure}

\textit{Noisy Gell-Mann operators}: For this case, as we did for qubit operators, we have to normalise this by dividing it with the quantity:
\begin{align*}
	\frac{I_{G'}}{I_{G}}&=\frac{\int d^3\xi e^{-\varepsilon ((1-\lambda_1)^{2}\xi^{2}_{1}+(1-\lambda_2)^{2}\xi^{2}_{2}+(1-\lambda_3)^{2}\xi^{2}_{3})} }{\int d^3\xi e^{-\varepsilon(\xi_1^2+\xi_2^2+\xi_3^2)}} \\
	&=\frac{1}{(1-\lambda_1)(1-\lambda_2)(1-\lambda_3)},
\end{align*}
where $I_G'$ is the integral of the scaled Gaussian and $I_G$ is the integral of the Gaussian function over the Fourier space. The negative volume of the Wigner distribution, thus, can be calculated as:
\begin{align}\label{n3noisynegqudit}
	\mathscr{W}_{neg}^{noisy}(\rho,\vec{A})=(1-\lambda_1)(1-\lambda_2)(1-\lambda_3)\mathscr{W}_{neg}(\rho,\vec{A}).
\end{align}
We see that for an arbitrary dimensional system(qudit) the negative volume of the Wigner function decreases with an increase in the noise vis-à-vis increase in the compatibility. We can numerically integrate the Eqn.(\ref{n3noisynegqudit}) and Eqn.(\ref{mubnegqutrit}) to see how the negative volume varies with an increase in the dimension of the system for a fixed value of noise. Fig.(\ref{fig:12}) shows that the negative volume decrease with an increase in the dimension of the system. In this case we have chosen $\lambda_1=\lambda_2=\lambda_3=\lambda$
\subsection{Two operators (n=2)}
\subsubsection{Gell-Mann operators}
If we choose our operators to be $A_k=\Lambda^{N}_{k},\ k=1,2$ then a general Wigner distribution can still be written as:
\begin{equation}\label{Quditn2wigidentorho}
    \mathscr{W}_{\rho}(\vec{a})=\frac{1}{N}(1+\vec{r} \cdot \vec{a})\mathscr{W}_{\mathbb{I}}(\vec{a}),
\end{equation}
where $r_k=0$ for $k>3$ and
\begin{equation}
\begin{gathered}
\Lambda_{1}^N=\begin{pmatrix}
0 & 1 & 0 & \cdot & \cdot & \cdot & 0\\
1 & 0 & 0 & \cdot & \cdot & \cdot & 0\\
\cdot & \cdot & \cdot & \cdot &   &   & \cdot\\
\cdot & \cdot & \cdot &   & \cdot &   & \cdot\\
\cdot & \cdot & \cdot &   &   & \cdot & \cdot\\
0 & 0 & 0 & \cdot & \cdot & \cdot & 0
\end{pmatrix},
\qquad
\Lambda_{2}^N=\begin{pmatrix}
0 & -i & 0 & \cdot & \cdot & \cdot & 0\\
i & 0 & 0 & \cdot & \cdot & \cdot & 0\\
\cdot & \cdot & \cdot & \cdot &   &   & \cdot\\
\cdot & \cdot & \cdot &   & \cdot &   & \cdot\\
\cdot & \cdot & \cdot &   &   & \cdot & \cdot\\
0 & 0 & 0 & \cdot & \cdot & \cdot & 0
\end{pmatrix},
\end{gathered}
\end{equation}
The expression for Fourier transform of the Wigner function for identity comes out to be $\widehat{\mathscr{W}_{\mathbb{I}}}(\vec{\xi})=(N-2)+2\cos{|\xi|}$ where $|\xi|=\sqrt{\xi_1^2+\xi_2^2}$. Then it follows that the Wigner function for identity comes out to be
\begin{align}\label{Quditn2mubidenwigner}
	\mathscr{W}_{\mathbb{I}}(\vec{a})&=\frac{1}{(2 \pi)^{2}} \int d^2 \xi e^{-i \vec{\xi} \cdot \vec{a}} ((N-2)+2\cos(|\xi|))e^{-\varepsilon \xi}\nonumber\\
	&\rightarrow\frac{-1}{\pi\left(1-|a|^{2}\right)^{3 / 2}}.
\end{align}
Note that, we still are working in the limit $\varepsilon\rightarrow0$. Using Eqn.(\ref{Quditn2wigidentorho}) and Eqn.(\ref{Quditn2mubidenwigner}) we get
\begin{equation}\label{Quditn2mubrhowigner}
    \mathscr{W}_{\rho}(\vec{a})=-\frac{(1+\vec{r} \cdot \vec{a})}{N\pi}\frac{1}{\left(1-|a|^{2}\right)^{3 / 2}}.
\end{equation}
\subsubsection{Noisy Gell-Mann Operators (Scaled Regularization)}
After adding the noise the operator set becomes $A_k=(1-\lambda_k)\Lambda^{N}_k+\lambda_k\mathbb{I},\ k=1,2$. The Fourier transform of Wigner function comes out to be $\widehat{\mathscr{W}}_{\mathbb{I}}^{noisy}(\vec{\xi})=e^{i (\xi_{1} \lambda_1 +\xi_{2} \lambda_2)}((N-2)+2\cos(|\xi'|)$ with $|\xi'|=\sqrt{(1-\lambda_1)^{2}\xi^{2}_{1}+(1-\lambda_2)^{2}\xi^{2}_{2}}$. The convoluted Wigner function for identity is then given by
\begin{align}\label{Quditn2noisyidenwigner}
	\mathscr{W}_{\mathbb{I}}(\vec{a})&=\frac{1}{(2 \pi)^{2}} \int d^2 \xi e^{-i \vec{\xi}' \cdot \vec{a}} (1+2\cos(|\xi'|))e^{-\varepsilon \xi'}\nonumber\\
	&\frac{-1}{\pi(1-\lambda_1)(1-\lambda_2)\left(1-|a'|^{2}\right)^{3 / 2}},
\end{align}
where $a'_k=\frac{a_k-\lambda_k}{1-\lambda_k}$. The total Wigner function turns out to be:
\begin{equation}\label{Quditn2mubrhowignersymm}
    \mathscr{W}_{\rho}(\vec{a}')=-\frac{(1+\vec{r} \cdot \vec{a}')}{N\pi(1-\lambda_1)(1-\lambda_2)}\frac{1}{\left(1-|a'|^{2}\right)^{3 / 2}}.
\end{equation}
\subsubsection{Comparing the negativity}
First, we will calculate the negative volume for the Wigner distribution of noiseless operators. We get
\begin{align}\label{mubnegn2qutrit}
	\mathscr{W}_{neg}(\rho,\vec{A})&= \int\frac{1}{2}\left(\mathscr{W}_{\rho}(\vec{a})-|\mathscr{W}_{\rho}(\vec{a})|\right)d^3a  \nonumber\\
	&= \int\frac{1}{2}\Bigg(-\frac{(1+\vec{r} \cdot \vec{a})}{N\pi}\frac{1}{\left(1-|a|^{2}\right)^{3 / 2}} \nonumber \\
	&-\Big|-\frac{(1+\vec{r} \cdot \vec{a})}{N\pi}\frac{1}{\left(1-|a|^{2}\right)^{3 / 2}}\Big|\Bigg)d^2a.
\end{align}
We see that it is inversely proportional to the dimension of the system N. We will see that the same thing follows for the noise case also.

\begin{figure}
    \centering
    \includegraphics[scale=0.7]{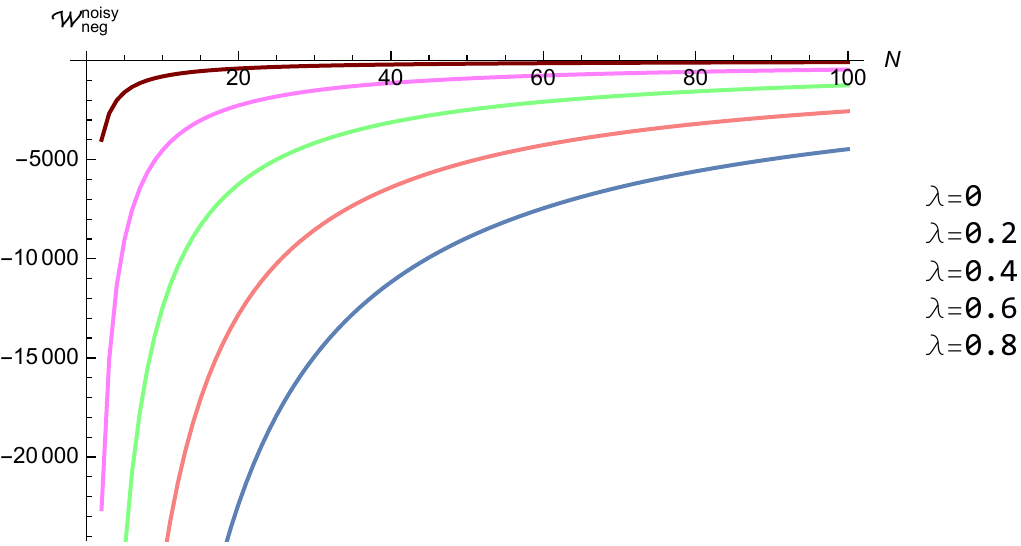}
    \caption{Variation of negativity of Wigner distribution function with the dimension of the operators N for different amounts of noise $\lambda$ under scaled Gaussian regularization}
    \label{fig:13}
\end{figure}

\textit{Noisy Gell-Mann operators}: By adding the noise the negative volume of the Wigner function for the qudit operator can be calculated in the same way as it was done for qubit operators. After dividing it by the normalizing factor
\begin{align*}
	\frac{I_{G'}}{I_{G}}&=\frac{\int d^2\xi e^{-\varepsilon\sqrt{ (1-\lambda_1)^{2}\xi^{2}_{1}+(1-\lambda_2)^{2}\xi^{2}_{2}} }}{\int d^2\xi e^{-\varepsilon\sqrt{\xi_1^2+\xi_2^2}}} \\
	&=\frac{1}{(1-\lambda_1)(1-\lambda_2)}.
\end{align*}
Then the expression comes out to be:
\begin{align}\label{noisyneg1qudit}
	\mathscr{W}_{neg}^{noisy}(\rho,\vec{A})=(1-\lambda_1)(1-\lambda_2)\mathscr{W}_{neg}^{mub}(\rho,\rho{A}).
\end{align}
This expression indicates that the negative volume of the Wigner function decreases with an increase in the noise injection for any arbitrary dimension. Fig.(\ref{fig:13}) can be generated by numerically integrating Eqn.(\ref{noisyneg1qudit}) and Eqn.(\ref{mubnegn2qutrit})for a particular value of the noise parameter (here we have $\lambda_1=\lambda_2=\lambda$. We observe that the negative volume of the Wigner distribution decreases with an increase in the system's dimension for any chosen noise value.\\

\section{Conclusion and outlook}\label{Sec8}
In this work, we have established a connection between the negativity of a quasiprobability distribution function, namely the Wigner function, and the incompatibility among the given set of observables. Both of these are very valuable resources in quantum information theory in general. Thereupon the treatment employed in this work can prove to be quite useful. Initially, we applied our treatment to the noisy eigenprojections of the qubit Pauli operators and observed that with an increase in the noise (decrease in the incompatibility) the negativity of the Wigner distribution decreases. We then applied the same formalism to the noisy qubit Pauli operators and obtained the same behavior. The negative volume maximizes for the maximally unbiased noiseless Pauli qubit operators. We then extended this treatment for higher dimensional qudit Gell-Mann operators. The same correlation between the negative volume of the Wigner distribution and incompatibility was observed there also.  Although our connection is qualitative up to some extent, we have a strong reason to believe that it provides a very good indication and comparison of the relative values of degrees of incompatibility among different observables using the negativity of the Wigner distribution as the index.

As discussed, with the increase in noise, and hence compatibility, the Wigner function starts becoming more and more positive but the use of the different regularising functions forgoes us from quantifying it. We can only comment on the relative amounts of the degree of incompatibility. Thus it will be worth finding out whether the regularising function can be kept unchanged during the whole treatment or, going one step further, even excluded to index the degree of incompatibility exactly through the negativity of the Wigner function.

It is also worth mentioning that the negative volume of the Wigner distribution function indicates towards being a measure of incompatibility for three observables. For two observables commutator is a known measure of incompatibility between them but up to our knowledge, there is no concrete measure of incompatibility for three and more observables. Thus studying the negativity of this Wigner distribution function definition is a very promising direction to look into. The only problem is that due to the high singularities in the Wigner function, it might be difficult to calculate it. One approach to handle this caveat is to convolute it with some smooth functions that provide symmetry to the integral which assists us in integrating it. As we pointed out in earlier sections this will render the analysis qualitative. To keep it quantitative to index the incompatibility through the negativity of the Wigner distribution function we have to opt for the numerical integration way. These aspects are very interesting and worth exploring in the future.

One more interesting direction to explore would be generalizing the definition of the Wigner function itself to other quasiprobability distributions. Also, in this paper, we are only dealing with static and finite-dimensional observables. We leave the extension to dynamic and infinite dimensional observables as a subject for future research.
\newpage
\bibliographystyle{apsrev4-2}
\bibliography{arxiv}

\begin{thebibliography}{17}%
\makeatletter
\providecommand \@ifxundefined [1]{%
 \@ifx{#1\undefined}
}%
\providecommand \@ifnum [1]{%
 \ifnum #1\expandafter \@firstoftwo
 \else \expandafter \@secondoftwo
 \fi
}%
\providecommand \@ifx [1]{%
 \ifx #1\expandafter \@firstoftwo
 \else \expandafter \@secondoftwo
 \fi
}%
\providecommand \natexlab [1]{#1}%
\providecommand \enquote  [1]{``#1''}%
\providecommand \bibnamefont  [1]{#1}%
\providecommand \bibfnamefont [1]{#1}%
\providecommand \citenamefont [1]{#1}%
\providecommand \href@noop [0]{\@secondoftwo}%
\providecommand \href [0]{\begingroup \@sanitize@url \@href}%
\providecommand \@href[1]{\@@startlink{#1}\@@href}%
\providecommand \@@href[1]{\endgroup#1\@@endlink}%
\providecommand \@sanitize@url [0]{\catcode `\\12\catcode `\$12\catcode
  `\&12\catcode `\#12\catcode `\^12\catcode `\_12\catcode `\%12\relax}%
\providecommand \@@startlink[1]{}%
\providecommand \@@endlink[0]{}%
\providecommand \url  [0]{\begingroup\@sanitize@url \@url }%
\providecommand \@url [1]{\endgroup\@href {#1}{\urlprefix }}%
\providecommand \urlprefix  [0]{URL }%
\providecommand \Eprint [0]{\href }%
\providecommand \doibase [0]{https://doi.org/}%
\providecommand \selectlanguage [0]{\@gobble}%
\providecommand \bibinfo  [0]{\@secondoftwo}%
\providecommand \bibfield  [0]{\@secondoftwo}%
\providecommand \translation [1]{[#1]}%
\providecommand \BibitemOpen [0]{}%
\providecommand \bibitemStop [0]{}%
\providecommand \bibitemNoStop [0]{.\EOS\space}%
\providecommand \EOS [0]{\spacefactor3000\relax}%
\providecommand \BibitemShut  [1]{\csname bibitem#1\endcsname}%
\let\auto@bib@innerbib\@empty
\bibitem [{\citenamefont {Heisenberg}(1925)}]{heisenberg}%
  \BibitemOpen
  \bibfield  {author} {\bibinfo {author} {\bibfnamefont {W.}~\bibnamefont
  {Heisenberg}},\ }\href {https://doi.org/10.1007/BF01328377} {\bibfield
  {journal} {\bibinfo  {journal} {Zeitschrift f{\"u}r Physik}\ }\textbf
  {\bibinfo {volume} {33}},\ \bibinfo {pages} {879} (\bibinfo {year}
  {1925})}\BibitemShut {NoStop}%
\bibitem [{\citenamefont {Bohr}\ \emph {et~al.}(1928)\citenamefont {Bohr} \emph
  {et~al.}}]{bohr1928quantum}%
  \BibitemOpen
  \bibfield  {author} {\bibinfo {author} {\bibfnamefont {N.}~\bibnamefont
  {Bohr}} \emph {et~al.},\ }\href@noop {} {\emph {\bibinfo {title} {The quantum
  postulate and the recent development of atomic theory}}},\ Vol.~\bibinfo
  {volume} {3}\ (\bibinfo  {publisher} {Printed in Great Britain by R. \& R.
  Clarke, Limited},\ \bibinfo {year} {1928})\BibitemShut {NoStop}%
\bibitem [{\citenamefont {Fine}(1982)}]{PhysRevLett.48.291}%
  \BibitemOpen
  \bibfield  {author} {\bibinfo {author} {\bibfnamefont {A.}~\bibnamefont
  {Fine}},\ }\href {https://doi.org/10.1103/PhysRevLett.48.291} {\bibfield
  {journal} {\bibinfo  {journal} {Phys. Rev. Lett.}\ }\textbf {\bibinfo
  {volume} {48}},\ \bibinfo {pages} {291} (\bibinfo {year} {1982})}\BibitemShut
  {NoStop}%
\bibitem [{\citenamefont {Masanes}\ \emph {et~al.}(2006)\citenamefont
  {Masanes}, \citenamefont {Acin},\ and\ \citenamefont
  {Gisin}}]{PhysRevA.73.012112}%
  \BibitemOpen
  \bibfield  {author} {\bibinfo {author} {\bibfnamefont {L.}~\bibnamefont
  {Masanes}}, \bibinfo {author} {\bibfnamefont {A.}~\bibnamefont {Acin}},\ and\
  \bibinfo {author} {\bibfnamefont {N.}~\bibnamefont {Gisin}},\ }\href
  {https://doi.org/10.1103/PhysRevA.73.012112} {\bibfield  {journal} {\bibinfo
  {journal} {Phys. Rev. A}\ }\textbf {\bibinfo {volume} {73}},\ \bibinfo
  {pages} {012112} (\bibinfo {year} {2006})}\BibitemShut {NoStop}%
\bibitem [{\citenamefont {Wolf}\ \emph {et~al.}(2009)\citenamefont {Wolf},
  \citenamefont {Perez-Garcia},\ and\ \citenamefont
  {Fernandez}}]{PhysRevLett.103.230402}%
  \BibitemOpen
  \bibfield  {author} {\bibinfo {author} {\bibfnamefont {M.~M.}\ \bibnamefont
  {Wolf}}, \bibinfo {author} {\bibfnamefont {D.}~\bibnamefont {Perez-Garcia}},\
  and\ \bibinfo {author} {\bibfnamefont {C.}~\bibnamefont {Fernandez}},\ }\href
  {https://doi.org/10.1103/PhysRevLett.103.230402} {\bibfield  {journal}
  {\bibinfo  {journal} {Phys. Rev. Lett.}\ }\textbf {\bibinfo {volume} {103}},\
  \bibinfo {pages} {230402} (\bibinfo {year} {2009})}\BibitemShut {NoStop}%
\bibitem [{\citenamefont {Quintino}\ \emph {et~al.}(2014)\citenamefont
  {Quintino}, \citenamefont {V\'ertesi},\ and\ \citenamefont
  {Brunner}}]{PhysRevLett.113.160402}%
  \BibitemOpen
  \bibfield  {author} {\bibinfo {author} {\bibfnamefont {M.~T.}\ \bibnamefont
  {Quintino}}, \bibinfo {author} {\bibfnamefont {T.}~\bibnamefont
  {V\'ertesi}},\ and\ \bibinfo {author} {\bibfnamefont {N.}~\bibnamefont
  {Brunner}},\ }\href {https://doi.org/10.1103/PhysRevLett.113.160402}
  {\bibfield  {journal} {\bibinfo  {journal} {Phys. Rev. Lett.}\ }\textbf
  {\bibinfo {volume} {113}},\ \bibinfo {pages} {160402} (\bibinfo {year}
  {2014})}\BibitemShut {NoStop}%
\bibitem [{\citenamefont {Uola}\ \emph {et~al.}(2014)\citenamefont {Uola},
  \citenamefont {Moroder},\ and\ \citenamefont
  {G\"uhne}}]{PhysRevLett.113.160403}%
  \BibitemOpen
  \bibfield  {author} {\bibinfo {author} {\bibfnamefont {R.}~\bibnamefont
  {Uola}}, \bibinfo {author} {\bibfnamefont {T.}~\bibnamefont {Moroder}},\ and\
  \bibinfo {author} {\bibfnamefont {O.}~\bibnamefont {G\"uhne}},\ }\href
  {https://doi.org/10.1103/PhysRevLett.113.160403} {\bibfield  {journal}
  {\bibinfo  {journal} {Phys. Rev. Lett.}\ }\textbf {\bibinfo {volume} {113}},\
  \bibinfo {pages} {160403} (\bibinfo {year} {2014})}\BibitemShut {NoStop}%
\bibitem [{\citenamefont {Uola}\ \emph {et~al.}(2015)\citenamefont {Uola},
  \citenamefont {Budroni}, \citenamefont {G\"uhne},\ and\ \citenamefont
  {Pellonp\"a\"a}}]{PhysRevLett.115.230402}%
  \BibitemOpen
  \bibfield  {author} {\bibinfo {author} {\bibfnamefont {R.}~\bibnamefont
  {Uola}}, \bibinfo {author} {\bibfnamefont {C.}~\bibnamefont {Budroni}},
  \bibinfo {author} {\bibfnamefont {O.}~\bibnamefont {G\"uhne}},\ and\ \bibinfo
  {author} {\bibfnamefont {J.-P.}\ \bibnamefont {Pellonp\"a\"a}},\ }\href
  {https://doi.org/10.1103/PhysRevLett.115.230402} {\bibfield  {journal}
  {\bibinfo  {journal} {Phys. Rev. Lett.}\ }\textbf {\bibinfo {volume} {115}},\
  \bibinfo {pages} {230402} (\bibinfo {year} {2015})}\BibitemShut {NoStop}%
\bibitem [{\citenamefont {Kochen}\ and\ \citenamefont
  {Specker}(1975)}]{kochen1975problem}%
  \BibitemOpen
  \bibfield  {author} {\bibinfo {author} {\bibfnamefont {S.}~\bibnamefont
  {Kochen}}\ and\ \bibinfo {author} {\bibfnamefont {E.~P.}\ \bibnamefont
  {Specker}},\ }in\ \href@noop {} {\emph {\bibinfo {booktitle} {The
  logico-algebraic approach to quantum mechanics}}}\ (\bibinfo  {publisher}
  {Springer},\ \bibinfo {year} {1975})\ pp.\ \bibinfo {pages}
  {293--328}\BibitemShut {NoStop}%
\bibitem [{\citenamefont {Xu}\ and\ \citenamefont
  {Cabello}(2019)}]{PhysRevA.99.020103}%
  \BibitemOpen
  \bibfield  {author} {\bibinfo {author} {\bibfnamefont {Z.-P.}\ \bibnamefont
  {Xu}}\ and\ \bibinfo {author} {\bibfnamefont {A.}~\bibnamefont {Cabello}},\
  }\href {https://doi.org/10.1103/PhysRevA.99.020103} {\bibfield  {journal}
  {\bibinfo  {journal} {Phys. Rev. A}\ }\textbf {\bibinfo {volume} {99}},\
  \bibinfo {pages} {020103} (\bibinfo {year} {2019})}\BibitemShut {NoStop}%
\bibitem [{\citenamefont {Kenfack}\ and\ \citenamefont
  {{\.Z}yczkowski}(2004)}]{kenfack2004negativity}%
  \BibitemOpen
  \bibfield  {author} {\bibinfo {author} {\bibfnamefont {A.}~\bibnamefont
  {Kenfack}}\ and\ \bibinfo {author} {\bibfnamefont {K.}~\bibnamefont
  {{\.Z}yczkowski}},\ }\href@noop {} {\bibfield  {journal} {\bibinfo  {journal}
  {Journal of Optics B: Quantum and Semiclassical Optics}\ }\textbf {\bibinfo
  {volume} {6}},\ \bibinfo {pages} {396} (\bibinfo {year} {2004})}\BibitemShut
  {NoStop}%
\bibitem [{\citenamefont {Hudson}(1974)}]{HUDSON1974249}%
  \BibitemOpen
  \bibfield  {author} {\bibinfo {author} {\bibfnamefont {R.}~\bibnamefont
  {Hudson}},\ }\href
  {https://doi.org/https://doi.org/10.1016/0034-4877(74)90007-X} {\bibfield
  {journal} {\bibinfo  {journal} {Reports on Mathematical Physics}\ }\textbf
  {\bibinfo {volume} {6}},\ \bibinfo {pages} {249} (\bibinfo {year}
  {1974})}\BibitemShut {NoStop}%
\bibitem [{\citenamefont {Gross}(2006)}]{gross2006hudson}%
  \BibitemOpen
  \bibfield  {author} {\bibinfo {author} {\bibfnamefont {D.}~\bibnamefont
  {Gross}},\ }\href {https://aip.scitation.org/doi/pdf/10.1063/1.2393152}
  {\bibfield  {journal} {\bibinfo  {journal} {Journal of mathematical physics}\
  }\textbf {\bibinfo {volume} {47}},\ \bibinfo {pages} {122107} (\bibinfo
  {year} {2006})}\BibitemShut {NoStop}%
\bibitem [{\citenamefont {Spekkens}(2008)}]{PhysRevLett.101.020401}%
  \BibitemOpen
  \bibfield  {author} {\bibinfo {author} {\bibfnamefont {R.~W.}\ \bibnamefont
  {Spekkens}},\ }\href {https://doi.org/10.1103/PhysRevLett.101.020401}
  {\bibfield  {journal} {\bibinfo  {journal} {Phys. Rev. Lett.}\ }\textbf
  {\bibinfo {volume} {101}},\ \bibinfo {pages} {020401} (\bibinfo {year}
  {2008})}\BibitemShut {NoStop}%
\bibitem [{\citenamefont {Rahimi-Keshari}\ \emph {et~al.}(2021)\citenamefont
  {Rahimi-Keshari}, \citenamefont {Mehboudi}, \citenamefont {De~Santis},
  \citenamefont {Cavalcanti},\ and\ \citenamefont
  {Ac\'{\i}n}}]{PhysRevA.104.042212}%
  \BibitemOpen
  \bibfield  {author} {\bibinfo {author} {\bibfnamefont {S.}~\bibnamefont
  {Rahimi-Keshari}}, \bibinfo {author} {\bibfnamefont {M.}~\bibnamefont
  {Mehboudi}}, \bibinfo {author} {\bibfnamefont {D.}~\bibnamefont {De~Santis}},
  \bibinfo {author} {\bibfnamefont {D.}~\bibnamefont {Cavalcanti}},\ and\
  \bibinfo {author} {\bibfnamefont {A.}~\bibnamefont {Ac\'{\i}n}},\ }\href
  {https://doi.org/10.1103/PhysRevA.104.042212} {\bibfield  {journal} {\bibinfo
   {journal} {Phys. Rev. A}\ }\textbf {\bibinfo {volume} {104}},\ \bibinfo
  {pages} {042212} (\bibinfo {year} {2021})}\BibitemShut {NoStop}%
\bibitem [{\citenamefont {Schwonnek}\ and\ \citenamefont
  {Werner}(2020)}]{schwonnek2020wigner}%
  \BibitemOpen
  \bibfield  {author} {\bibinfo {author} {\bibfnamefont {R.}~\bibnamefont
  {Schwonnek}}\ and\ \bibinfo {author} {\bibfnamefont {R.~F.}\ \bibnamefont
  {Werner}},\ }\href {https://aip.scitation.org/doi/abs/10.1063/1.5140632}
  {\bibfield  {journal} {\bibinfo  {journal} {Journal of Mathematical Physics}\
  }\textbf {\bibinfo {volume} {61}},\ \bibinfo {pages} {082103} (\bibinfo
  {year} {2020})}\BibitemShut {NoStop}%
\bibitem [{\citenamefont {Heinosaari}\ \emph {et~al.}(2016)\citenamefont
  {Heinosaari}, \citenamefont {Miyadera},\ and\ \citenamefont
  {Ziman}}]{Heinosaari_2016}%
  \BibitemOpen
  \bibfield  {author} {\bibinfo {author} {\bibfnamefont {T.}~\bibnamefont
  {Heinosaari}}, \bibinfo {author} {\bibfnamefont {T.}~\bibnamefont
  {Miyadera}},\ and\ \bibinfo {author} {\bibfnamefont {M.}~\bibnamefont
  {Ziman}},\ }\href {https://doi.org/10.1088/1751-8113/49/12/123001} {\bibfield
   {journal} {\bibinfo  {journal} {Journal of Physics A: Mathematical and
  Theoretical}\ }\textbf {\bibinfo {volume} {49}},\ \bibinfo {pages} {123001}
  (\bibinfo {year} {2016})}\BibitemShut {NoStop}%
\end{thebibliography}%
\end{document}